\documentclass[aps,reprint,floatfix]{revtex4-1}
\usepackage{amsmath,amssymb,graphicx}
\usepackage{hhline,slashbox,bbm}

\makeatletter
\DeclareRobustCommand{\cev}[1]{%
  \mathpalette\do@cev{#1}%
}
\newcommand{\do@cev}[2]{%
  \fix@cev{#1}{+}%
  \reflectbox{$\m@th#1\vec{\reflectbox{$\fix@cev{#1}{-}\m@th#1#2\fix@cev{#1}{+}$}}$}%
  \fix@cev{#1}{-}%
}
\newcommand{\fix@cev}[2]{%
  \ifx#1\displaystyle
  \mkern#23mu
  \else
  \ifx#1\textstyle
  \mkern#23mu
  \else
  \ifx#1\scriptstyle
  \mkern#22mu
  \else
  \mkern#22mu
  \fi
  \fi
  \fi
}

\makeatother

\newcounter{theoremcounter}
\stepcounter{theoremcounter}

\newtheorem{corollary}{Corollary}

\newenvironment{definition}[1][Definition]{\begin{trivlist}
\item[\hskip \labelsep {\bfseries #1}]}{\end{trivlist}}

\newcommand{\bs}{\boldsymbol}
\newcommand{\bra}[1]{\left\langle #1\right|}
\newcommand{\ket}[1]{\left|#1\right\rangle}

\newcommand{\expval}[3]{\left\langle #1\middle|#2\middle|#3\right\rangle}
\newcommand{\ketbra}[2]{\ket{#1}\bra{#2}}
\newcommand{\partder}[2]{\frac{\partial #1}{\partial #2}}
\newcommand{\der}[2]{\frac{\mbox d #1}{\mbox d #2}}
\DeclareMathOperator{\Tr}{Tr}

\makeatletter
\newcommand{\vast}{\bBigg@{4}}
\newcommand{\Vast}{\bBigg@{5}}
\makeatother

\begin{document}

\title{Discrete Wigner Formalism for Qubits and Non-Contextuality of Clifford Gates on Qubit Stabilizer States}
\author{Lucas Kocia}
\affiliation{Department of Physics, Tufts University, Medford, Massachusetts 02155, U.S.A.}
\author{Peter Love}
\affiliation{Department of Physics, Tufts University, Medford, Massachusetts 02155, U.S.A.}
\begin{abstract}
  We show that qubit stabilizer states can be represented by non-negative quasi-probability distributions associated with a Wigner-Weyl-Moyal formalism where Clifford gates are positive state-independent maps. This is accomplished by generalizing the Wigner-Weyl-Moyal formalism to three generators instead of two---producing an exterior, or Grassmann, algebra---which results in Clifford group gates for qubits that act as a permutation on the finite Weyl phase space points naturally associated with stabilizer states. As a result, a non-negative probability distribution can be associated with each stabilizer state's three-generator Wigner function, and these distributions evolve deterministically to one another under Clifford gates. This corresponds to a hidden variable theory that is non-contextual and local for qubit Clifford gates while Clifford (Pauli) measurements have a context-dependent representation. Equivalently, we show that qubit Clifford gates can be expressed as propagators within the three-generator Wigner-Weyl-Moyal formalism whose semiclassical expansion is truncated at order \(\hbar^0\) with a finite number of terms. The \(T\)-gate, which extends the Clifford gate set to one capable of universal quantum computation, require a semiclassical expansion of the propagator to order \(\hbar^1\). We compare this approach to previous quasi-probability descriptions of qubits that relied on the two-generator Wigner-Weyl-Moyal formalism and find that the two-generator Weyl symbols of stabilizer states result in a description of evolution under Clifford gates that is state-dependent, in contrast to the three-generator formalism. We have thus extended Wigner non-negative quasi-probability distributions from the odd \(d\)-dimensional case to \(d=2\) qubits, which describe the non-contextuality of Clifford gates and contextuality of Pauli measurements on qubit stabilizer states.
\end{abstract}
\maketitle

\section{Introduction}
\label{sec:intro}

Contextuality~\cite{Kochen67,Redhead87,Mermin93} is a necessary resource for universal quantum computation~\cite{Howard14}. In general, the existence of a positive quasiprobability representation is a notion of classicality that is equivalent to non-contextuality~\cite{Spekkens08,Ferrie08,Ferrie09,Ferrie11}. As a result of work by Wootters~\cite{Wootters87}, Gross~\cite{Gross06}, Veitch \emph{et al}.~\cite{Veitch12,Veitch13}, Mari \emph{et al}.~\cite{Mari12}, and Howard \emph{et al}.~\cite{Howard14}, it has been established that non-contextuality is equivalent to the non-negativity of the discrete Wigner functions, and Weyl symbols, of the associated states and operators involved~\cite{Spekkens08}. In particular, it is possible to formulate a discrete two-parameter or two-generator Wigner function for odd \(d\)-dimensional qudit~\cite{Gross06} and rebit~\cite{Delfosse15} stabilizer states that are non-negative, along with positive covariant maps for the associated Weyl symbols of the Clifford gates. For odd \(d\), quantum gates and states that are non-contextual can be efficiently simulated on classical computers~\cite{Mari12,Howard14,Kocia17}.

However, it is impossible to define a non-negative two-generator Wigner function for qubit stabilizer states and positive covariant two-generator Clifford Weyl symbols~\cite{Gross08,Veitch12,Mari12,Zhu15,Zhu16}. This is true despite the fact that it has long been known that qubit stabilizer states and Clifford operations can be simulated efficiently by the Gottesmann-Knill theorem~\cite{Aaronson04}, and that contextuality is necessary for attaining quantum universality in qubit circuits with some additional postulates~\cite{Raussendorf15}. This begs the question: is contextuality only equivalent to the non-negativity of associated discrete Wigner functions for \(d\)-dimensional qudits with odd \(d\)?

Here we answer this question. We find that non-contextuality is equivalent to non-negativity in the (appropriate) associated discrete Wigner function for qubits (\(d=2\)). The issue preventing such a result in previous efforts was the use of only two generators to define a discrete Wigner function. By instead employing three generators, and thereby defining an exterior (or Grassmann) algebra, we show that the Wigner-Weyl-Moyal formalism, first developed by Berezin~\cite{Berezin77}, results in discrete Wigner functions that are non-negative for stabilizer states, and Weyl symbols that are state-independent positive maps for Clifford gates. This is related to the fact that Clifford gates in any odd prime power dimension are unitary two-designs, while multi\emph{qubit} Clifford gates are also unitary three-designs~\cite{Zhu15}. The necessity of using three bases for qubits was also found by Wallman \emph{et al}.~\cite{Wallman12}, from a very different approach to the Wigner-Weyl-Moyal formalism, to characterize one qubit non-contexutality.

We further show that the Weyl symbol for qubit unitary gates can be formulated in terms of a traditional path integral expansion in powers of \(\hbar\) and find that Clifford gates can be fully described by a single term consisting of the truncated path integral at order \(\hbar^0\). On the other hand, the \(T\)-gate, which extends the Clifford gates to a universal quantum gate set, requires the full path integral up to order \(\hbar^1\). This agrees with results found for odd \(d\)-dimensional qudits~\cite{Kocia16,Dax17}. The Weyl symbols of Pauli measurements, with which Clifford gates complete the set of Clifford operations, are shown to have the possibility of being contextual, when more than one qubit is involved in the system.

Finally, we show how the Weyl symbols of stabilizer states in the two-generator formalism relate to their three-generator counterparts. Using a map from the two-generator algebra stabilizer states to the three-generator algebra stabilizer states, we explain their propagation rules under Clifford gates. This yields the Aaronson-Gottesman tableau algorithm, just as in the odd-\(d\) case~\cite{Kocia17}. In this two-generator representation we find that Clifford gates must be defined state-dependently, whereas in the three-generator representation, evolution is state independent. As a result, we argue that the two-generator formalism forms a non-local and contextual hidden variable theory for Clifford gates on qubit stabilizer states from the perspective of preparation contextuality. This is in contrast to the three-generator representation, which forms a local and non-contextual hidden variable theory.

We begin by first offering motivation in Section~\ref{sec:motivation} for why formulating a discrete Wigner function for qudits with only two generators is necessarily restricted to odd \(d\) if it is to be associated with the usual Wigner-Weyl-Moyal formalism. We introduce some fundamentals of the Grassmann algebra in Section~\ref{sec:fundamentals}. This allows us to introduce the Wigner-Weyl-Moyal formalism with three generators in Section~\ref{sec:WWMformalism}, and show that within this framework Clifford gates and stabilizer states are positive state-independent maps and non-negative states, respectively, for \(d=2\). We return to discrete two-generator Wigner functions in Section~\ref{sec:twogenerators}, which includes Wootter's original discrete formulation for qubits~\cite{Wootters87}, and relate it to the three-generator algebra. We explain how the Aaronson-Gottesman tableau algorithm for qubit stabilizer state propagation under Clifford gates is equivalent to such a two-generator Wigner function, as we argued in recent work~\cite{Kocia17}. Finally, in Section~\ref{sec:measurement}, we discuss how the contextuality of Pauli measurements is manifest in the three-generator Wigner-Weyl-Moyal formalism.

\section{Motivation for Using Three Generators}
\label{sec:motivation}

Most prior formulations of a discrete Wigner function can be expressed as a discretization of the continuous two-generator Wigner-Weyl-Moyal formalism~\cite{Almeida98,Rivas99,Ruzzi05} to odd \(d\). The Wigner formalism replaces operators and states by their corresponding Weyl symbols (defined below). Therein, the usual conjugate momentum (\(p\)) and position (\(q\)) degrees of freedom are replaced with the ``center'' and ``chord'' degrees of freedom. This turns out to be very useful in the discrete case where \(\hat p\) and \(\hat q\) no longer form a Lie algebra in which their Lie product (commutator) is a scalar. Instead, the generator of the corresponding one-parameter Lie subgroups, \(e^{i \theta \hat p}\) and \(e^{i \phi \hat q}\), obey a simple Weyl relation (which can be interpreted as a weaker group commutation relation~\cite{Kocia16}) and these define the corresponding translations and reflections that chords and centers parametrize, respectively.

Prior formulations of such a discrete Wigner-Weyl-Moyal formalism have generally relied on expanding the state \(\rho \in L(\mathbb C^{d^n})\) in a basis of operators labelled by the points of a \((\mathbb Z/d\mathbb Z)^n \times (\mathbb Z/d\mathbb Z)^n\) grid~\cite{Veitch12}, for \(n\) \(d\)-dimensional qudits. They depend on a discretization of the following Weyl-Heisenberg operators, which are also called generalized translation operators in semiclassics~\cite{Almeida98}:
\begin{equation}
  \label{eq:weylheisenberg}
  \hat T(\bs \lambda_p, \bs \lambda_q) = \exp \left(-\frac{i}{2\hbar} \bs \lambda_p \cdot \bs \lambda_q\right) \hat Z^{\bs \lambda_p} \hat X^{\bs \lambda_q}.
\end{equation}
The set \(\hat T\) are Hilbert-Schmidt orthogonal.
\(\hat Z\) and \(\hat X\) generate a Lie group and correspond to the ``boost'' operator:
\begin{equation}
  \hat Z^{\delta p} \ket{q'} = e^{\frac{i}{\hbar} \hat q \delta p} \ket{q'} = e^{\frac{i}{\hbar} q' \delta p} \ket{q'},
\end{equation}
and the ``shift'' operator:
\begin{equation}
  \hat X^{\delta q} \ket{q'} = e^{-\frac{i}{\hbar} \hat p \delta q} \ket{q'} = \ket{q' + \delta q},
\end{equation}
which satisfy the Weyl relation:
\begin{equation}
  \label{eq:contWeylrelation}
  \hat Z \hat X =  e^{\frac{i}{\hbar}} \hat X \hat Z.
\end{equation}
From Eq.~\ref{eq:contWeylrelation} and Eq.~\ref{eq:weylheisenberg}, it follows that
\begin{equation}
  \hat T^\dagger(\bs \lambda_p, \bs \lambda_q) = \hat T(-\bs \lambda_p, -\bs \lambda_q).
\end{equation}
The translation operator defines the characteristic function of an operator \(\hat \rho\):
\begin{equation}
  \label{eq:twogencharfunction}
  \rho_\lambda(\bs \lambda_p, \bs \lambda_q) = \Tr \left( \hat T^\dagger(\bs \lambda_p, \bs \lambda_q) \hat \rho \right).
\end{equation}
This is the chord representation of \(\hat \rho\). We define \(\hat R(x)\) as the symplectic Fourier transform of \(\hat T(\lambda)\):
\begin{eqnarray}
  \label{eq:contreflection}
  &&\hat R(\bs x_p, \bs x_q) = \left(2 \pi \hbar\right)^{-n} \int^\infty_{-\infty} \mbox{d} \bs \lambda e^{\frac{i}{\hbar} {\bs \lambda}^T \bs{\mathcal J} {\bs x} } \hat T(\bs \lambda),
\end{eqnarray}
where
\begin{equation}
\bs{\mathcal J} =  \left( \begin{array}{cc} 0 & -\mathbb{I}_{n}\\ \mathbb{I}_{n} & 0 \end{array}\right),
\end{equation}
for \(\mathbb{I}_n\) the \(n\)-dimensional identity.

These \(\hat R\)-operators are Hermitian, Hilbert-Schmidt orthogonal, self-inverse and therefore also unitary:
\begin{equation}
  \hat R^{-1}(\bs x) = \hat R^\dagger(\bs x) = \hat R(\bs x).
\end{equation}
With this in hand, the Weyl symbol of operator \(\hat \rho\) can be expressed as the coefficient of the density matrix expanded in the basis of states \(\hat R(\bs x_p, \bs x_q)\):
\begin{equation}
  {\rho}_x(\bs x_p, \bs x_q) = \Tr \left( \hat {R}^\dagger(\bs x_p, \bs x_q) \hat \rho \right).
  \label{eq:twogenweylsymbol}
\end{equation}
\(\bs x \equiv (\bs x_p, \bs x_q) \in \mathbb{R}^{2n}\) are called ``centers'' or Weyl phase space points. If \(\hat \rho\) is a state, \(\rho_x\) is the corresponding Wigner function.

Restricting this to finite \(d\)-dimensional systems involves setting \(\hbar = d/2\pi\), and enforcing periodic boundary conditions~\cite{Rivas99}. The points \((\bs \lambda_p, \bs \lambda_q)\) and \((\bs x_p, \bs x_q)\) become elements in \((\mathbb{Z} / d \mathbb{Z})^{2n}\) and form a discrete ``web'' or Weyl ``grid''. The generalized translation operator becomes
\begin{equation}
\hat {T}(\bs \lambda_p, \bs \lambda_q) = \omega^{-\bs \lambda_p \cdot \bs \lambda_q (d+1)/2} \hat {Z}^{ \bs \lambda_p} \hat {X}^{ \bs \lambda_q}.
\end{equation}
where \(\omega\equiv \exp 2 \pi i/d\) and \((d+1)/2\) is equivalent to \(1/2\) in mod odd-\(d\) arithmetic.
In this way, it can be seen that the generalized translation operator plays a fundamental role in the definition of the Wigner function. In particular, it defines a Lie algebra with two generators, \(\hat p\) and \(\hat q\).

Unfortunately, even with different definitions for \(\omega\), the translation operator forms a subgoup of \(SU(d)\) only for \(d\) odd, i.e. for \(d=2\), it is in \(U(2)\), not \(SU(2)\)~\cite{Bengtsson17}. This can be seen by evaluating
\begin{equation}
  \det \hat T(1,0) = \det \hat T(0,1) = \det \hat T(1,1) = (-1)^{d+1},
\end{equation}
which is only equal to \(1\) for odd \(d\).
This also manifests itself in \(2d\)- instead of \(d\)-periodicity for some elements:
\begin{equation}
  \left( \hat T(1,0) \hat T(0,1) \right)^2 = \left(\hat Z \hat X\right)^2 = (i \hat Y)^2 = -1.
\end{equation}
As a result, it becomes impossible by the approach detailed by Eqs.~\ref{eq:twogencharfunction}-\ref{eq:twogenweylsymbol}, to find the results we expect of Wigner functions for the even \(d\) case. In particular, we would like our Wigner-Weyl-Moyal formalism to have the following properties:
\begin{enumerate}
\item Stabilizer states are the discrete analogues of Gaussians and so have non-negative Wigner functions.
\item Clifford operators have underlying harmonic Hamiltonians and are positive state-independent maps that can be treated by a path integral truncated at order \(\hbar^0\).
\end{enumerate}

There are an infinite number of possible formulations for a discrete two-generator Wigner-Weyl-Moyal formalism, which are related to each other by unitary transformations. It has been shown that none of them satisfy the above characteristics~\cite{Gross08,Veitch12,Mari12,Raussendorf15,Delfosse15,Zhu15,Zhu16}.

Here we show that all these different discrete Wigner formulations eventually run into trouble for even \(d\) because they inevitably must keep track of another degree of freedom, and they accomplish this by using both positive and negative numbers in either a Clifford gate Weyl symbol, or a stabilizer state Wigner function.

For qubits the problem can be traced to the fact that the Weyl-Heisenberg group is a subgroup of \(U(2)\) and not \(SU(2)\). The remedy to put the Weyl-Heisenberg group back into \(SU(2)\) is to change the algebra for the degrees of freedom to the Grassmann algebra. The resultant Weyl-Wigner-Moyal formalism can then be made to satisfy conditions \(1-2\) above with the addition of another degree of freedom. We will find that this third degree of freedom, which we will refer to as `\(r\)' to complement the usual `\(p\)' and `\(q\)' degrees of freedom, will accomplish this task without resorting to negativity.

\section{Some Grassmann Fundamentals}
\label{sec:fundamentals}

A discrete system with \(d=2\) (spin-\(1/2\)) has no classical mechanical counterpart. However, one can invoke canonical quantization in reverse, and determine a classical mechanical system which yields spin-\(1/2\) when canonically quantized. This problem was solved by Berezin in 1977~\cite{Berezin77} in which he identified the Grassmann algebra with three generators as the appropriate ``pseudo-''classical system corresponding to spin-\(1/2\) under canonical quantization. Berezin showed that this formalism is interpretable in terms of Weyl symbols. However, it appears that the semiclassical Wigner-Weyl-Moyal formalism of Grassmann numbers has not been developed, where translations and reflections, as in Eq.~\ref{eq:weylheisenberg} and Eq.~\ref{eq:contreflection}, are identified~\footnote{There has been much work on the Wigner-Weyl-Moyal formalism for fermions that is somewhat related.}. Here, we develop this semiclassical formalism and derive the propagator in powers of \(\hbar\).

As we shall show, the Grassman algebra with three generators provides not only the classical system corresponding to spin \(1/2\) under canonical quantization, but also a subtheory of spin-\(1/2\) which is the familiar qubit stabilizer formalism with Clifford operators.

An exterior---or Grassmann---algebra is an associative algebra that contains a vector space such that the square of any vector space element is zero. More formally, the Grassmann algebra over the vector space \(V\) over the field \(K\) is defined as the quotient algebra of the tensor algebra by the two-sided ideal \(I\) generated by all elements of the form \(x \otimes x\) for \(x \in V\).

Let \(\xi_p\), \(\xi_q\) and \(\xi_r\) be three real generators of a Grassmann algebra \(\mathcal G_3\). Hence,
\begin{equation}
\xi_j \xi_k + \xi_k \xi_j \equiv \{\xi_j, \xi_k\} = 0,  \quad \text{for} \, j,k \in \{1,2,3\},
\end{equation}
where we can identify \(\xi_p \equiv \xi_1\), \(\xi_q \equiv \xi_2\) and \(\xi_r \equiv \xi_3\). Any element \(g \in \mathcal G_3\) may be represented as a finite sum of homogeneous monomials consisting of the three generators:
\begin{eqnarray}
  \label{eq:Grassmannscalar}
  g(\bs \xi) &=& g_0 \\
  \label{eq:Grassmannvector}
             && + \left(g_p \xi_p + g_q \xi_q + g_r \xi_r\right)\\
  \label{eq:Grassmannaxialvector}
             && + \left( g_{pq} \xi_p \xi_q + g_{qr} \xi_q \xi_r + g_{pr} \xi_p \xi_r\right.\\
             && \quad \left.+ g_{qp} \xi_q \xi_p + g_{rq} \xi_r \xi_q + g_{rp} \xi_r \xi_p\right) \nonumber\\
  \label{eq:Grassmanndeterminant}
             && + \left(g_{pqr} \xi_p \xi_q \xi_r + g_{prq} \xi_p \xi_r \xi_q + g_{rpq} \xi_r \xi_p \xi_q\right.\\
             && \quad \left.+ g_{rqp} \xi_r \xi_q \xi_p + g_{qrp} \xi_q \xi_r \xi_p + g_{qpr} \xi_q \xi_p \xi_r\right), \nonumber
\end{eqnarray}
where there is no implicit sum on \(p\), \(q\) and \(r\) and \(\bs \xi \equiv (\xi_p, \xi_q, \xi_r)\)~\cite{Berezin77}. Notice that the above equation is written in a manner to make the antisymmetry present explicit, i.e. \(g_{pq} \xi_p \xi_q = - g_{pq} \xi_q \xi_p\) and so the coefficients \(g_{pq}\) and \(g_{qp}\) can be exchanged under sign inversion. We write \(\bs \xi\) as the argument of \(g\) since we will relate \(g(\bs \xi)\) later to the Weyl symbol of the corresponding operator \(\hat g\). Any such element can be written as a linear combination of grades---each grade denotes monomials of the same degree---thereby forming a graded algebra. In particular, every element consists of a linear combination of a scalar (line~\ref{eq:Grassmannscalar}), vector (line~\ref{eq:Grassmannvector}), axial vector (line~\ref{eq:Grassmannaxialvector}), and a pseudoscalar (line~\ref{eq:Grassmanndeterminant}) grade. 

Next, we define an analog of complex conjugation by the following involution:
\begin{eqnarray}
  \label{eq:Grassmanncomplexconj}
  (g^*)^* &=& g,\\
  (\alpha g)^* &=& \alpha^* g^*,\nonumber
\end{eqnarray}
where \(\alpha\) is a complex number. If we define an element \(g\) as real if \(g^* = g\) (and an algebra as real if all its elements are real: \(\xi_k^* = \xi_k\)), then it follows that
\begin{equation}
  (g_1 g_2)^* = g_2^* g_1^*.
\end{equation}
This definition ensures that \(g g^*\) is real (since \((g g^*)^* = g g^*\)). Since we consider the generators \(\xi_k\) to be real, it follows that \(\xi_k^* = \xi_k\) by definition.

We can define derivatives as the following linear operators in \(\mathcal G_3\):
\begin{eqnarray}
  \frac{\vec \partial}{\partial \xi_l} \xi_{k_1} \cdots \xi_{k_\nu} &=& \delta_{k_1 l} \xi_{k_2} \cdots \xi_{k_\nu} - \delta_{k_2 l} \xi_{k_1} \xi_{k_3} \cdots \xi_{k_\nu} \nonumber \\
  \label{eq:leftderivative}
  && + \ldots + (-1)^\nu \delta_{k_\nu l} \xi_{k_1} \cdots \xi_{k_{\nu-1}},
\end{eqnarray}
and
\begin{eqnarray}
  \xi_{k_1} \cdots \xi_{k_\nu} \frac{\cev \partial}{\partial \xi_l} &=& \delta_{k_\nu l} \xi_{k_1} \cdots \xi_{k_{\nu-1}} - \delta_{k_{\nu-1} l} \xi_{k_1}  \cdots \xi_{k_{\nu-2}} \xi_{k_\nu} \nonumber \\ 
  \label{eq:rightderivative}
  && + \ldots + (-1)^\nu \delta_{k_1 l} \xi_{k_2}\cdots \xi_{k_{\nu-1}}.
\end{eqnarray}
The operator \(\vec \partial/\partial \xi_l\) is the left derivative and \(\cev \partial/\partial \xi_l\) is the right derivative. Examining Eq.~\ref{eq:leftderivative} and \ref{eq:rightderivative}, we can see that the left derivative of a monomial can be found by permuting \(\xi_l\) to the left and then dropping it and vice-versa for the right derivative.

With derivatives thus defined, we can develop the analog of the definite single integral (over the whole support of a variable):
\begin{equation}
  \label{eq:Grassmannintegral1}
  \int 1 \mbox d \xi_l = 0,
\end{equation}
and
\begin{equation}
  \label{eq:Grassmannintegral2}
  \int \xi_l \mbox d \xi_l = 1.
\end{equation}
These can be generalized to multiple integration:
\begin{equation}
  \label{eq:Grassmannintegral3}
  \int \xi_{k_1} \cdots \xi_{k_\nu} \mbox d \xi_\nu \cdots \mbox d \xi_1 = \epsilon_{k_1 \cdots k_\nu},
\end{equation}
and so
\begin{equation}
  \label{eq:Grassmannintegral4}
  \int g(\xi) \mbox d \xi_3 \mbox d \xi_2 \mbox d \xi_1 = \sum_{k_1,k_2,k_3= 1}^{3}\epsilon_{k_1 k_2 k_3} g_{k_1 k_2 k_3},
\end{equation}
where \(\epsilon_{k_1 \cdots k_\nu}\) is the Levi-Cevita tensor.

In this algebra, the Fourier transform \(\mathcal F\) can be described as a linear mapping \(\mathcal G_3 \rightarrow \tilde {\mathcal G}_3\) for the Grassmann algebras \(\mathcal G_3\) and \(\tilde{\mathcal G}_3\) with generators \(\xi_k\) and \(\rho_k\), \(k=1,2,3\) respectively, defined by:
\begin{equation}
  \label{eq:GrassmannFouriertransform}
  g(\bs \xi) = \mathcal F \left(\tilde g(\bs \rho)\right) = \int e^{ i \sum_k \xi_k \rho_k } \tilde g(\rho) \mbox d \rho_3 \mbox d \rho_2 \mbox d \rho_1,
\end{equation}
and its inverse
\begin{equation}
  \label{eq:GrassmanninverseFouriertransform}
  \tilde g(\bs \rho) = \mathcal F^{-1}\left( g(\bs \xi)\right) = i \int e^{ - i \sum_k \xi_k \rho_k } g(\xi) \mbox d \xi_3 \mbox d \xi_2 \mbox d \xi_1.
\end{equation}
Using the properties of the Grassmann elements and the integrals (Eq.~\ref{eq:Grassmannintegral4}), we find
\begin{widetext}
\begin{eqnarray}
  \tilde g(\bs \rho) &=& i \int \left(1 - i \sum_k \xi_k \rho_k - \frac{1}{2} \sum_{k,l} \xi_k \rho_k \xi_l \rho_l + \frac{i}{6} \sum_{k,l,m} \xi_k \rho_k \xi_l \rho_l \xi_m \rho_m\right) \nonumber\\
  \label{eq:oddgworkedout}
  && \quad \times \left[g_0 + \left(g_p \xi_p + g_q \xi_q + g_r \xi_r\right) + \left( g_{pq} \xi_p \xi_q + g_{qr} \xi_q \xi_r + g_{pr} \xi_p \xi_r + g_{qp} \xi_q \xi_p + g_{rq} \xi_r \xi_q + g_{rp} \xi_r \xi_p \right) \right.\\
  && \quad \quad \left.+ \left( g_{pqr} \xi_p \xi_q \xi_r + g_{prq} \xi_p \xi_r \xi_q + g_{rpq} \xi_r \xi_p \xi_q + g_{rqp} \xi_r \xi_q \xi_p + g_{qrp} \xi_q \xi_r \xi_p + g_{qpr} \xi_q \xi_p \xi_r\right) \right] \mbox d \xi_r \mbox d \xi_q \mbox d \xi_p \nonumber\\
  &=& i \left(g_{pqr} - g_{prq} + g_{rpq} - g_{qpr} + g_{qrp}\right) +\left(- \rho_p g_{qr} + \rho_q g_{pr} - \rho_r g_{pq} + \rho_p g_{rq} - \rho_q g_{rp} + \rho_r g_{qp}\right) \nonumber\\
  && -i \left(\rho_p \rho_r g_q -i \rho_p \rho_q g_r +i \rho_r \rho_q g_p \right) + \rho_p \rho_q \rho_r g_0. \nonumber
\end{eqnarray}
\end{widetext}
Substituting this equation for \(\tilde g(\bs \rho)\) back into Eq.~\ref{eq:GrassmannFouriertransform} produces Eq.~\ref{eq:Grassmannscalar}-\ref{eq:Grassmanndeterminant} for \(g(\bs \xi)\).

From this exercise it is clear that the vector-space grade is dual to the axial vectors and the scalars are dual to the pseudoscalars under the Fourier transform.

In this way, the Fourier transformation takes even monomials to odd monomials and vice-versa. Therefore, we can restrict \(g(\bs \xi)\) in Eq.~\ref{eq:Grassmannscalar}-\ref{eq:Grassmanndeterminant} to only include even or odd monomials. Following the natural notation above, we will call the even representation \(g(\bs \xi)\) and the odd one \(\tilde g(\bs \rho)\) where
\begin{eqnarray}
  \label{eq:centerrep}
  g(\bs \xi) &=& g_0 + \left( g_{pq} \xi_p \xi_q + g_{qr} \xi_q \xi_r + g_{pr} \xi_p \xi_r\right.\\
  && \quad \quad \left. + g_{qp} \xi_q \xi_p + g_{rq} \xi_r \xi_q + g_{rp} \xi_r \xi_p\right), \nonumber
\end{eqnarray}
and
\begin{eqnarray}
  \tilde g(\bs \rho) &=& \tilde g_p \rho_p + \tilde g_q \rho_q + \tilde g_r \rho_r \nonumber\\
  \label{eq:chordrep}
                     && + \left(\tilde g_{pqr} \rho_p \rho_q \rho_r + \tilde g_{prq} \rho_p \rho_r \rho_q + \tilde g_{rpq} \rho_r \rho_p \rho_q\right.\\
                     && \quad \left. + \tilde g_{rqp} \rho_r \rho_q \rho_p + \tilde g_{qrp} \rho_q \rho_r \rho_p + \tilde g_{qpr} \rho_q \rho_p \rho_r \right), \nonumber
\end{eqnarray}
such that
\begin{equation}
  g(\bs \xi) = \mathcal F \left( \tilde g(\bs \rho) \right).
\end{equation}

The even Grassmann polynomial representation \(g(\bs\xi)\) is thus dual to the odd Grassmann polynomial representation \(\tilde g(\bs \rho)\) by the Fourier transform \(\mathcal F\). In particular, from Eq.~\ref{eq:oddgworkedout} it can be shown that the terms defining \(g(\bs \xi)\) and \(\tilde g(\bs \rho)\) in Eq.~\ref{eq:centerrep} and Eq.~\ref{eq:chordrep} are related by:\begin{eqnarray}
  \tilde g_p &=& (g_{rq} - g_{qr}),\\
  \tilde g_q &=& (g_{pr} - g_{rp}),\\
                                                                                                                                                                                                                                                                                                                                                                                                                                 \tilde g_r &=& (g_{qp} - g_{pq}),\\
                                                                                                                                                                                                                                                                                                                                                                                                                                 \begin{array}{c}
                                                                                                                                                                                                                                                                                                                                                                                                                                    \tilde g_{pqr} - \tilde g_{prq} + \tilde g_{rpq}\\
                                                                                                                                                                                                                                                                                                                                                                                                                                    - \tilde g_{rqp} + \tilde g_{qrp} - \tilde g_{qpr}
                                                                                                                                                                                                                                                                                                                                                                                                                                 \end{array}&=& g_0.
\end{eqnarray}

We note that the usual plane wave identity of the Dirac delta function,
\begin{equation}
  \label{eq:Fourierdelta}
  i \int \exp\left( i \sum_k \rho_k \xi_k \right) \text{d}^3 \rho = \xi_p \xi_q \xi_r \equiv \delta(\xi_p) \delta(\xi_q) \delta(\xi_r),
\end{equation}
holds since,
\begin{equation}
  \int g(\bs \xi) \xi_p \xi_q \xi_r \text{d}^3 \xi = g(0),
\end{equation}
which follows from Eq.~\ref{eq:Grassmannintegral4}.

We can further define a Gaussian integral, which will prove useful when we evaluate the path integral for a harmonic Hamiltonian later~\cite{Berezin77}:
\begin{equation}
  \label{eq:GrassmannGaussianintegral}
  \int \exp \left( \sum a_{jk} \xi_j \xi_k \right) \mbox d \xi_3 \mbox d \xi_2 \mbox d \xi_3 = \sqrt{\det | 2 a_{jk} |},
\end{equation}
where \(a_{jk} = -a_{kj}\). Notice that the resultant determinant is in the numerator instead of the denominator, unlike in the usual Gaussian integral over \(\mathbb R\) or \(\mathbb C\).

As a final point, we note that the three real generators can be treated as classical canonical variables:
\begin{equation}
  i \sum_{j} \left(\xi_k \frac{\cev \partial}{\partial \xi_j}\right) \left(\frac{\vec \partial}{\partial \xi_j} \xi_l\right) = \{\xi_k, \xi_l\}_{\text{P.B.}} = i \delta_{kl},
\end{equation}
where ``P. B.'' stands for the Poisson bracket. Therefore, their evolution can be found from their Poisson bracket with a Hamiltonian \(H\):
\begin{equation}
  \label{eq:eomGrassmann}
  \frac{\mbox d}{\mbox d t} \xi_k = \{H,\xi_k\}_{\text{P.B}} = i H \frac{\cev \partial}{\partial \xi_k}.
\end{equation}

We will see later that measurement outcomes and expectation values are contained in the scalar grade, or the dual pseudoscalar grade. Unitary operators are contained in the axial vector grade or the dual vector grade. Projectors and density matrices will be found to be linear combinations of both the scalar and axial vector grades, or the dual pseudoscalar and vector grades.

\section{Quantum Wigner-Weyl-Moyal Formalism with Three Generators}
\label{sec:WWMformalism}

To quantize our algebra, we replace the Poisson brackets for the canonical variables by the anti-commutator multiplied by \(-i/\hbar\)~\cite{Berezin77}:
\begin{equation}
  \{\xi_k, \xi_l\}_{\text{P.B.}} \rightarrow \{\hat \xi_k, \hat \xi_l\} = \hbar \delta_{kl}.
\end{equation}
Renormalizing, we get the Clifford algebra with the three generators:
\begin{equation}
  \label{eq:GrassmannPauli}
  \hat \xi_k = \sqrt{\frac{\hbar}{2}} \hat \sigma_k,
\end{equation}
which obey
\begin{equation}
  \{\hat \sigma_k, \hat \sigma_l\} = 2 \delta_{kl}.
\end{equation}
We will set \(\hbar=2\) from now on to avoid writing this scaling factor everywhere. These \(\hat \sigma_k\) are the Pauli operators. As such, we know that the generators obey
\begin{equation}
  \label{eq:Paulimatrixrelation}
  i \hat \xi_1 \hat \xi_2 \hat \xi_3 = 1.
\end{equation}

Any operator \(\hat g\) may be written in terms of the operators \(\hat \xi\) as follows, where products of \(\hat \xi\) denote matrix products of the corresponding Pauli matrices:
\begin{eqnarray}
  \hat g &=& g_0 \\
         && + \left(g_p \hat \xi_p + g_q \hat \xi_q + g_r \hat \xi_r\right) \nonumber\\
         && + \left( g_{pq} \hat \xi_p \hat \xi_q + g_{qr} \hat \xi_q \hat \xi_r + g_{pr} \hat \xi_p \hat \xi_r\right. \nonumber\\
         && \quad \left.+ g_{qp} \hat \xi_q \hat \xi_p + g_{rq} \hat \xi_r \hat \xi_q + g_{rp} \hat \xi_r \hat \xi_p\right) \nonumber\\
         && + \left(g_{pqr} \hat \xi_p \hat \xi_q \hat \xi_r + g_{prq} \hat \xi_p \hat \xi_r \hat \xi_q + g_{rpq} \hat \xi_r \hat \xi_p \hat \xi_q\right. \nonumber\\
         && \quad \left.+ g_{rqp} \hat \xi_r \hat \xi_q \hat \xi_p + g_{qrp} \hat \xi_q \hat \xi_r \hat \xi_p + g_{qpr} \hat \xi_q \hat \xi_p \hat \xi_r\right). \nonumber
 \end{eqnarray}
As a result of Eq.~\ref{eq:Paulimatrixrelation}, this decomposition is unique only if even or odd terms are included~\cite{Berezin77}. Therefore, \(\hat g\) has two equivalent decompositions: even and odd. Note that there was a similar relationship for the even and odd representations \(g(\bs \xi)\), but the two were related by Fourier transform instead of being formally equal.

Motivated by this analogy, we define the Weyl symbol for the operator \(\hat g\) as equal to the \(g(\bs \xi)\) in Eqs.~\ref{eq:Grassmannscalar}-\ref{eq:Grassmanndeterminant}.

It follows that any operator can be expressed as a linear superposition of
\begin{equation}
  \label{eq:Grassmanntranslationop}
  \hat T(\bs \rho) = \exp \left( i \sum_k \hat \xi_k \rho_k\right)
\end{equation}
such that
\begin{equation}
  \hat g = \int \hat T(\bs \rho) \tilde g(\bs \rho) \text d^3 \rho.
\end{equation}

We point out that the integral symbol here means that these operators are, in a generalized way, labelled by a continuous set of Grassmann algebra elements instead of a finite set, as is the case in the discretized two-generator Wigner-Weyl-Moyal formalism. Nevertheless, we will soon see that this does not prevent us from associating a finite set of ``Weyl phase space points'' to stabilizer propagation.

Berezin identified the \(\hat T(\bs \rho)\) operator as a translation operator since from the anti-commutator~\cite{Berezin77} one can find,
\begin{equation}
  \hat T(\bs \rho') \hat T(\bs \rho'') = \exp\left( \sum_k \rho'_k \rho''_k \right) \hat T(\bs \rho' + \bs \rho'')
\end{equation}
and
\begin{equation}
\Tr \hat T(\bs \rho) = 2\left( 1 + i \rho_1 \rho_2 \rho_3 \right).
\end{equation}

Though \(\hat T(\bs \rho)\), as defined in Eq.~\ref{eq:Grassmanntranslationop}, is an operator, it does not live in the Hilbert space; \(\hat T(\bs \rho)\) does not take states in the Hilbert space to other states in the Hilbert space. However, Berezin showed that it can be used to go between Hilbert space and the Grassmann algebra \(\mathcal G_3\):
\begin{equation}
  \label{eq:translationtrace}
  \Tr\left(\hat T(-\bs \rho) \hat g \right) = 2 \left( i \tilde g(\bs \rho) + g\left( i \bs \rho \right) \right).
\end{equation}

Since \(g(\bs \xi)\) and \(\tilde g(\bs \rho)\) have opposite parities, this equation provides the Weyl symbol of \(\hat g\) if you choose \(g(\bs \xi)\) or \(\tilde g(\bs \rho)\) to contain only even terms, and its dual to only contain odd terms.

Whereas \(\hat g_1 \hat g_2 = \hat g\) can be found by the multiplication rules of the Pauli matrices, the Weyl symbol is able to reproduce the same algebraic structure with the Grassmann elements via an integral:
\begin{eqnarray}
  \hat g &\equiv& \hat g_1 \hat g_2 \nonumber\\
         &=& \int \hat T(\bs \rho') \tilde g_1(\bs \rho') \text d^3 \rho' \int \hat T(\bs \rho'') \tilde g_2(\bs \rho'') \text d^3 \rho'' \nonumber\\
         &=& \int \exp\left( \sum_k \rho'_k \rho''_k \right) \hat T(\bs \rho' + \bs \rho'') \tilde g_1(\bs \rho') \tilde g_2(\bs \rho'') \text d^3 \rho' \text d^3 \rho'' \nonumber\\
         &=& \int \exp\left(\sum_k \rho'_k \rho''_k\right) \hat T(\bs \rho' + \bs \rho'') \\
         &&\quad \times \left(\int \exp\left(-i \sum_k \xi'_k \rho'_k \right) g_1(\bs \xi') \text d^3 \xi' \right)\nonumber\\
         &&\quad \times \left( \int \exp\left(-i \sum_k \xi''_k \rho''_k \right) g_2(\bs \xi'') \text d^3 \xi'' \right) \text d^3 \rho' \text d^3 \rho'' \nonumber,
\end{eqnarray}
With Eq.~\ref{eq:GrassmannFouriertransform} for the Weyl symbol in terms of an integral over an exponential, we can find the Weyl symbol:
\begin{eqnarray}
  g(\bs \xi) &=& \int e^{ \sum_k \rho'_k \rho''_k + i\sum_k \xi_k (\rho' + \rho'')_k - i \sum_k (\xi'_k\rho'_k+ \xi''_k \rho''_k)} \nonumber\\
  && \quad \times g_1(\bs \xi') g_2(\bs \xi'') \text d^3 \xi' \text d^3 \xi'' \text d^3 \rho' \text d^3 \rho'' \nonumber\\
  &=& \int e^{i \sum_k \left( \xi_k \rho''_k - \xi''_k \rho''_k \right)} g_1(\bs \xi') g_2(\bs \xi'') \\
  && \quad \times \delta\left( \bs \rho'' + i (\bs \xi- \bs \xi') \right)  \text d^3 \xi' \text d^3 \xi'' \text d^3 \rho'' \nonumber\\
  \label{eq:productoftwoweylsymbols}
  &=& \int g_1(\bs \xi') g_2(\bs \xi'') e^{ \Delta_3(\bs \xi, \bs \xi', \bs \xi'')} \text d^3 \xi' \text d^3 \xi'', \nonumber
\end{eqnarray}
where
\begin{equation}
  \Delta_3(\bs \xi, \bs \xi_1, \bs \xi_2) = \sum_k \left(\xi'_k \xi''_k + \xi''_k \xi_k + \xi_k \xi'_k\right).
\end{equation}
This can be extended to the product of three operators:
\begin{eqnarray}
  (g_1 g_2 g_3)(\bs \xi) &=& \int g_1(\bs \xi') g_2(\bs \xi''+\bs \xi'-\bs \xi) g_3(\bs \xi'') \nonumber\\
  && \qquad \quad \times e^{ \Delta_3(\bs \xi, \bs \xi', \bs \xi'', \bs \xi''')} \text d^3 \xi' \text d^3 \xi'',
\end{eqnarray}
and so on. This identity will prove useful later when we discuss the propagation of density operators \(\hat \rho \rightarrow \hat U \hat \rho \hat U^\dagger\).

We can define the dual to the translation \(\hat T\) operator:
\begin{equation}
  \hat R(\bs \xi) = \int \exp(-i \sum_k \xi_k \rho'_k) \hat T(\bs \rho') \text d^3 \rho'.
\end{equation}

It follows that
\begin{eqnarray}
  \hat R(\bs \xi) \hat T(\bs \rho) &=& \int \exp(-i \sum_k \xi_k \rho'_k) \hat T(\bs \rho') \hat T(\bs \rho) \text d^3 \rho' \nonumber\\
                                &=& \int \exp \sum_k \left(-i \xi_k \rho'_k + \rho'_k \rho_k \right) \hat T(\bs \rho' + \bs \rho) \text d^3 \rho' \nonumber\\
                                &=& \int \exp \left(i \sum_k \rho'_k \left(\xi - i \rho \right)_k \right) \hat T(\bs \rho' + \bs \rho) \text d^3 \rho' \nonumber\\
                                &=& \int \exp \left( i \sum_k \left(\eta - \rho \right)_k \left( \xi - i \rho \right)_k \right) \hat T(\bs \eta) \text d^3 \eta \nonumber\\
                                &=& \exp \left(-i \sum_k \rho_k \left(\xi - i \rho \right)_k \right) \hat R \left(\bs \xi - i \bs \rho \right) \\
                                &=& \exp \left(-i \sum_k \rho_k \xi_k \right) \hat R \left( \bs \xi - i \bs \rho \right). \nonumber
\end{eqnarray}

Similarly,
\begin{eqnarray}
  \hat T(\bs \rho) \hat R(\bs \xi) &=& \hat T(\bs \rho) \int \exp(-i \sum_k \xi_k \rho'_k) \hat T(\bs \rho') \text d^3 \rho' \nonumber\\
                                &=& \int \exp \left(-i \sum_k \xi_k \rho'_k + \sum_k \rho_k \rho'_k \right) \hat T(\bs \rho + \bs \rho') \text d^3 \rho' \nonumber\\
                                &=& \int \exp \left( i \sum_k \rho'_k \left(\xi + i \rho \right)_k \right) \hat T(\bs \rho + \bs \rho') \text d^3 \rho' \\
                                &=& \int \exp \left( i \sum_k \left( \eta - \rho \right)_k \left( \xi + i \rho \right)_k \right) \hat T(\bs \eta) \text d^3 \eta \nonumber\\
                                &=& \exp \left(-i \sum_k \rho_k \left(\xi + i \rho \right)_k \right) \hat R \left( \bs \xi + i \bs \rho \right) \nonumber\\
                                &=& \exp \left(-i \sum_k \rho_k \xi_k \right) \hat R \left( \bs \xi + i \bs \rho \right). \nonumber
\end{eqnarray}

Finally, we can use this last result to find
\begin{eqnarray}
  &&\hat R(\bs \xi'') \hat R(\bs \xi') \nonumber\\
  &=& \int \exp(-i \sum_k \xi''_k \rho_k) \hat T(\bs \rho) \hat R(\bs \xi') \text d^3 \rho \nonumber\\
  &=& \int \exp(-i \sum_k \xi''_k \rho_k - i \sum_k \rho_k \xi'_k) \hat R \left(\bs \xi' + i \bs \rho \right) \text d^3 \rho \nonumber\\
  &=& \int \exp \left( i \sum_k (\eta - \xi')_k (\xi'' - \xi')_k \right) \hat R(\bs \eta) \text d^3 \rho \\
  &=& \exp \left(i \sum_k \xi'_k (\xi'' - \xi')_k \right) \hat T\left(\bs \xi'' - \bs \xi' \right) \nonumber\\
  &=& \exp \left(i \sum_k \xi'_k \xi''_k \right) \hat T \left(\bs \xi'' - \bs \xi'\right), \nonumber
\end{eqnarray}
where we also used the inverse Fourier transform:
\begin{equation}
  \hat T(\bs \rho) = \int \exp \left(i \sum_k \xi_k \rho_k \right) \hat R(\bs \xi) \text d^3 \xi.
\end{equation}
This shows that \(\hat R\) is self-inverse and so
\begin{equation}
  \hat R^2(\bs \xi) = 1.
\end{equation}

Therefore, it can be seen that \(\hat R(\bs \xi)\) corresponds to a general reflection (actually an inversion) operator just as in the two-generator formalism. Like the \(\hat T\)-operator, this is not a Hilbert space operator. However, taking the trace we see that it can be used to go between Hilbert space and \(\mathcal G_3\):
\begin{eqnarray}
  \label{eq:reflectiontrace}
  &&\Tr(\hat R(\bs \xi) \hat g) \nonumber\\
&=& \Tr \int \exp \left(i \sum_k \xi_k \rho_k \right) \hat T(\bs \rho) \hat g \text d^3 \rho \\
                           &=& \int \exp \left( i \sum_k \xi_k \rho_k \right) 2 \left[ i \tilde g(-\bs \rho) + g\left(-i \bs \rho \right)\right] \nonumber\\
                           &=& 2 \left[ i g(-\bs \xi) + \tilde g\left(i \bs \xi\right)\right]. \nonumber
\end{eqnarray}

Again, this equation provides the Weyl symbol \(g(\bs \xi)\) if you choose \(g(-\bs \xi)\) to be even and \(\tilde g \left( i \bs \xi \right)\) to be odd. Notice that this differs from the two-generator Wigner-Weyl-Moyal formalism where the Weyl symbol of an operator \(\hat \rho\) is found by just its trace with the reflection operator.

In this paper, we will define our Weyl symbol as the even Grassmann polynomial. Instead of pulling out the even terms of Eq.~\ref{eq:translationtrace} or Eq.~\ref{eq:reflectiontrace} to construct \(g(\xi)\), an easier way to find the Weyl symbol is to find the Pauli matrix representation of the operator, and then dequantize using Eq.~\ref{eq:GrassmannPauli}. In this way, we can find that the Weyl symbol for any one-qubit pure state is
\begin{equation}
  g(\bs \xi) = \frac{1}{2} \left(1 + \left( \alpha i \xi_r \xi_q + \beta i \xi_p \xi_q + \gamma i \xi_p \xi_r \right) \right),
\end{equation}
where \(\alpha^2 + \beta^2 + \gamma^2 = 1\), for \(\alpha, \, \beta, \, \gamma \in \mathbb R\).

We can identify \(\xi_{q_\pm} \equiv \frac{1}{2} \left(1 \mp i \xi_p \xi_r \right)\) with the one-qubit computational basis states.

In the usual two-generator Wigner formalism, integrating over one of the generators produces a marginal probability over the other. For instance, for the Wigner function of a state \(\hat \rho\):
\begin{equation}
  \int^\infty_{-\infty} \rho_x(x_p,x_q) \text d x_p = \expval{q}{\hat \rho}{q}.
\end{equation}

Similarly, the expectation value of operator \(\hat O\) of such a state is:
\begin{equation}
  (2 \pi \hbar)^{-1} \int^\infty_{-\infty} \int^\infty_{-\infty} \rho_x(x_p,x_q) O_x(x_p,x_q) \text d x_p \text d x_q = \Tr \left(\hat O \hat \rho \right).
\end{equation}
where \(O_x(x_p, x_q)\) is the Weyl symbol of the operator \(\hat O\).

Unlike its two-generator analog, the three-generator Weyl symbol cannot generally produce scalar values after partial traces; it is a map to \(\mathcal G_3\) after all, not \(\mathbb R\). To produce a real value, a three-generator Weyl symbol must traced over all of its three degrees of freedom. Taking such a full trace of \(g(\xi)\), multiplied by odd Weyl symbols of operators representing observables, produces expectation values. Marginals can be obtained as a special case of expectation values.

For instance, we can find that the state \(\hat \rho' = \ketbra{1}{1} = \frac{1}{2}(1 - i \hat \xi_p \hat \xi_r)\) and so has the corresponding Weyl symbol \(\rho'(\xi) = \frac{1}{2}(1 - i \xi_p \xi_r)\). Using the Grassmann integral equations (Eqs.~\ref{eq:Grassmannintegral1}-\ref{eq:Grassmannintegral4}), it is easy to see that taking the trace with the odd Weyl symbols of the computational basis states, \(q_{\pm}(\xi) = \frac{1}{2}(i\xi_p \xi_r \xi_q \pm \xi_q)\), produces
\begin{equation}
  2 i \int \rho'(\xi) q_{\pm}(\xi) \text d \xi_r \text d \xi_q \text d \xi_p = \begin{cases}\left|\Psi(0)\right|^2 = 0 & \text{for \(-\),}\\ \left|\Psi(1)\right|^2 = 1 & \text{for \(+\)}.\end{cases}
\end{equation}
These results together provide the marginal \(|\Psi(q)|^2\). In general the expectation values of the projectors onto the eigenstates of an observable simply give the marginal distribution of that observable.

Furthermore, taking the trace with the odd Weyl symbol of non-projective operators produces the usual expectation values as well. For instance, taking the expectation value of the odd Weyl symbol of \(\hat q\), \(q(\xi) = \xi_q\), with \(\rho'(\xi)\) produces:
\begin{equation}
  \Tr \rho'(\bs \xi) = 2 i \int \rho(\xi) \xi_q  \text d \xi_r \text d \xi_p \text d \xi_q = 1,
\end{equation}
which is equal to \(\Tr\left(\hat \rho' \hat \sigma_z\right)\) as expected.

Like in the usual Wigner formalism, a trace corresponds to an integral over all of phase space. Taking the trace of \(\rho'(\bs \xi)\) with its odd symbol \(\tilde \rho'(\bs \xi)\) shows that it is a valid normalized state:
\begin{equation}
  2 i \int \rho'(\bs \xi) \frac{1}{2}(-\xi_p \xi_q \xi_r + \xi_q) \text d \xi_r \text d \xi_q \text d \xi_p = 1.
\end{equation}

From this perspective, the odd representation \(\tilde \rho(\xi)\) is the characteristic function of the discrete Wigner function expressed by the even representation; it is related to the even representation, by the Fourier transform, though neither are a true (quasi-)probability distribution without incorporating integration over dual Weyl symbols. In this way, compared to the two-generator Wigner-Weyl-Moyal formalism, the odd representation can be related to the chord representation, while the even representation is similar to the center representation. This is further supported by noting that the even monomials forms a subalgebra of \(\mathcal G_3\) and the odd monomials do not. Another difference compared to the two-generator Wigner-Weyl-Moyal formalism, is that the three generator ``center function'', \(g(\bs \xi)\), of operator \(\hat g\) is found by taking either the trace of \(\hat g\) with the translation or reflection operators (see Eq.~\ref{eq:translationtrace} and Eq.~\ref{eq:reflectiontrace}) and pulling out the even terms, instead of just the reflection operator (Eq.~\ref{eq:twogenweylsymbol}).

When constructing the Wigner-Weyl-Moyal formalism with the two generators \(p\) and \(q\) instead, which are quantized by setting their commutator instead of anticommutator equal to \(i \hbar \delta_{kl}\), it is easy to show that the associated operators cannot be bounded -- they must be supported on an infinite Hilbert space. As a result, the Hilbert space must by ``discretized'' by invoking periodic boundary conditions and setting \(\hbar\) appropriately. However, this ``breaks'' the commutator which is no longer a scalar. On the other hand, constructing the Weyl formalism with three generators, as we did here by quantizing \(\xi_1\), \(\xi_2\), \(\xi_3\) when we set their anticommutator equal to \(i \hbar \delta_{kl}\), produces associated operators that \emph{can} be bounded --- they can be supported on a finite Hilbert space. Therefore, there is no need to further ``discretize'' the Hilbert space by invoking periodic boundary conditions etc., as is necessary in the two-generator case.

Perhaps the most disconcerting difference between the two- and three-generator Weyl symbols is that whereas the two-generator symbols are maps from \((\mathbb Z/ d \mathbb Z)^n\times(\mathbb Z/ d \mathbb Z)^n\) to \(\mathbb R\), the three-generator symbols map from \(\mathcal G_3\) to \(\mathcal G_3\), or equivalently, from \((V^n,V^n,V^n)\) (since the argument of \(g(\bs \xi)\) is \(\bs \xi \equiv (\xi_p^n, \xi_q^n, \xi_r^n)\)) to \(\mathcal G_3\). This may appear strange at first glance, but the Grassmann elements that make up \(\mathcal G_3\) can be given a simple representation as matrices, which can serve to demystify their nature. This is given in the Appendix.

\section{Classicality of Clifford Gates on Stabilizer States}
\label{sec:classicalityofcliffops}

For odd \(d\) dimensions, positive maps and non-negative (discrete) Wigner functions can be associated with classical gates and states that are not capable of quantum speed-up~\cite{Veitch12}. This can also be formalized by finding that their propagator is complete when truncated at order \(\hbar^0\), whereas an operation that makes use of quantum resources requires at least order \(\hbar^1\)~\cite{Kocia16}. In this section we obtain the same results for \(d=2\) with the three-generator formalism using the Grassmann algebra.

\subsection{Propagator \(\hbar\) Expansion}
\label{sec:vanvleckprefactor}

Since we've restricted our Weyl symbol to only contain even terms and we want the Hamiltonian to be real, a Hamiltonian for a single qubit must have the Weyl symbol \(H = -\frac{i}{2} \sum_{k,l,m} \epsilon_{klm} b_k \xi_l \xi_m\) up to a constant, for real coefficients \(b_k\). Since \(\xi_k\) are real, it follows that \(H^* = H\) using the analog to complex conjugation defined in Eq.~\ref{eq:Grassmanncomplexconj}.

Given such a real quadratic Hamiltonian \(H\), we can construct a unitary matrix \(\bs{\mathcal E} = \exp i H \frac{\cev \partial^2}{\partial \bs \xi^2}\) such that \(\bs \xi' = \bs{\mathcal E} \bs \xi\). The Cayley parametrization relates this unitary matrix \(\bs{\mathcal E}\) to an antisymmetric matrix \(\bs A\):
\begin{equation}
  \bs{\mathcal E} = (\bs A - i \bs I)(\bs A + i \bs I)^{-1}
\end{equation}
and so
\begin{eqnarray}
  \bs{\mathcal E} &=&\exp \left( i H \frac{\cev \partial^2}{\partial \bs \xi^2} t \right) \nonumber\\
  \label{eq:antisymmetriccayleyparam}
             &=&\exp \left(\sum_k \epsilon_{klm} b_k t\right)\\
             &=& \left[\bs I+ \sum_k \epsilon_{klm} \frac{b_k}{b} \tan \left(\frac{b}{2} t\right)\right] \nonumber\\
             && \times \left[\bs I- \sum_n \epsilon_{nop} \frac{b_n}{b} \tan \left(\frac{b}{2} t\right)\right]^{-1}, \nonumber
\end{eqnarray}
where \(b = |\bs b|\).

This parametrization in terms of an antisymmetric matrix is useful since we can identify \(\sum_k \epsilon_{klm}\frac{b_k}{b}\tan\left(\frac{1}{2} b t\right)\) with the quadratic part of the generating action:
\begin{equation}
  S(\bs \xi; t) = \epsilon_{klm}\frac{b_k}{b}\tan\left(\frac{1}{2} b t\right) \xi_l \xi_m,
\end{equation}
and follow the usual approach set by the traditional two-generator Wigner-Weyl-Moyal formalism~\cite{Almeida98}. Since we have chosen to represent our three-generator Weyl symbols by even polynomials, this is the full action up to a constant.

We can now proceed to construct the Weyl symbol of the semiclassical propagator
\begin{equation}
  U(\xi; t) = \mathcal N \exp (i S(\bs \xi; t)/\hbar),
\end{equation}
and solve for \(\mathcal N\) by enforcing the Weyl symbol of \(\hat U(t) \hat U^\dagger(t) = \hat I\) using Eq.~\ref{eq:productoftwoweylsymbols}:
\begin{eqnarray}
  1(\xi) &=& \int U(\bs \xi_1; t)  U(\bs \xi_2; t) e^{\frac{\hbar}{2} \Delta_3(\xi, \xi_1, \xi_2)} \\
         &=& \int e^{\frac{i}{\hbar} B_{kl} \xi_k \xi_l + \frac{i}{\hbar} B_{kl} \xi_k \xi_l + \frac{\hbar}{2} (\bs \xi_1 + \bs \xi) (\bs \xi_2 + \bs \xi)} \text d^n \bs \xi_1 \text d^n \bs \xi_2 \nonumber\\
         &=& \int e^{\frac{i}{\hbar} B_{kl} \xi_k \xi_l + \frac{i}{\hbar} B_{kl} \xi_k \xi_l + \frac{\hbar}{2} \bs \xi_1 \bs \xi_2} \text d^n \bs \xi_1 \text d^n \bs \xi_2, \nonumber
\end{eqnarray}
where \(B_{lm} \equiv \epsilon_{klm} \frac{b_k}{b} \tan(bt/2)\).

From the Gaussian integral identity (Eq.~\ref{eq:GrassmannGaussianintegral}),
\begin{equation}
  \implies \frac{1}{|\mathcal N|^2} = \left[ \det(\mathbb B - \tilde{\mathbb I})\right]^{1/2},
\end{equation}
where
\begin{equation}
  \mathbb B = \left(\begin{array}{cc} \bs B & \bs 0\\ \bs 0& \bs B \end{array}\right),
\end{equation}
and
\begin{equation}
  \tilde{\mathbb I} = \left(\begin{array}{cc} \bs 0 & \bs I\\ \bs I& \bs 0 \end{array}\right).
\end{equation}
Therefore,
\begin{equation}
  \mathcal N = \left[ \det(\mathbb B - \tilde{\mathbb I})\right]^{-1/4},
\end{equation}
up to a phase.

\begin{equation}
  \left[ \det(\mathbb B - \tilde{\mathbb I})\right]^{-\frac{1}{4}} = \left[ \det( \bs I - \bs B^2 ) \right]^{-\frac{1}{4}} = \left[ \det( \bs I \pm \bs B ) \right]^{-\frac{1}{2}}.
\end{equation}

Using the Cayley parameterization given by Eq.~\ref{eq:antisymmetriccayleyparam}, and the fact that \(\det \bs U(t) = 1\) \(\forall t\),
\begin{equation}
  \mathcal N = [\det(\bs I \pm \bs B)]^{-\frac{1}{2}} = 2 [\det(\bs I + \bs U)]^{\frac{1}{2}},
\end{equation}
up to a phase.

Hence,
\begin{eqnarray}
  U(\bs \xi, t) &=& \left[\det\left(\bs I + S \frac{\cev \partial^2 }{\partial \bs \xi^2}\right)\right]^{-\frac{1}{2}} \exp\left(\frac{i}{\hbar} S(\bs \xi;t)\right) \nonumber\\
  \label{eq:weylprop}
  &=& 2 [\det(\bs I + \bs U)]^{\frac{1}{2}} \exp\left(\frac{i}{\hbar} S(\bs \xi;t)\right).
\end{eqnarray}
Notice that the prefactor is in the numerator instead of the denominator, as in the usual two-generator Weyl symbol.

Substituting in our harmonic Hamiltonian and generating action, we find
\begin{equation}
  U(\bs \xi, t) = \cos\left(\frac{1}{2} b t\right) \exp \left[-\frac{2i}{\hbar} \epsilon_{klm}\frac{b_k}{b}\tan\left(\frac{1}{2} b t\right) \xi_l \xi_m \right],
\end{equation}
in agreement with Berezin's Weyl symbol of the propagator (Eq.~\(2.36\) in~\cite{Berezin77}), which he found through the full Feynman path integral.

Note that the semiclassical expression for the propagator in Eq.~\ref{eq:weylprop}, as in all such semiclassical treatments similar to the primitive Van Vleck-Morette-Gutzwiller propagator~\cite{Van28,Morette51,Gutzwiller67}, must only be correct up to \(\mathcal O (\hbar)\). However, here there are no higher order \(\hbar\) corrections and this expression is exact. In other words, the semiclassical propagator (defined as the propagator treated up to order \(\hbar\)) is equivalent to full quantum propagator for a qubit. Indeed, its associated operator is well-known:
\begin{equation}
  \hat U(t) = \exp\left[-\frac{i}{2} \bs b \cdot \hat{ \bs \sigma} t \right],
\end{equation}
(or, equivalently, Eq.~\ref{eq:qubitprop} shown below). This is not the case for \(d>2\), as has been shown~\cite{Kocia16,Dax17}, where higher than \(\hbar^1\) corrections generally exist, and a sum over more than one classical trajectory must be taken.

This can also be seen by noting that the equation for the evolution of the generators \(\xi_k\) can be written as~\cite{Almeida98}
\begin{equation}
  \frac{\text d}{\text d t} \xi_k \equiv \{ \{ H, \xi_k \} \} = \{H, \xi_k \}_{\text{P.B.}} + \mathcal O(\hbar^2),
\end{equation}
where \(\{\{\}\}\) is the Moyal bracket. The \(\mathcal O (\hbar^2)\) terms correspond to polynomials with power \(2\) and higher and so are zero~\cite{Gadella89}. It follows that the Grassmann evolution captures the dynamics up to order \(\hbar^1\) and is a full quantum treatment.

Since the propagation under every Hamiltonian can be treated at order \(\hbar^1\), and so its path integral consists of only one term, it follows that if the global phase can be neglected we can simply keep the absolute value of the propagator's prefactor. This corresponds to the van Vleck prefactor of the propagator:
\begin{equation}
  |U(\bs \xi, t)| =  \left| 2 (\det(\bs I + \bs U))^{\frac{1}{2}} \right| = \left| \cos \left( \frac{1}{2} b t\right) \right|,
\end{equation}
and is the \(\hbar^0\) part of the propagator.

If we can keep track of the \(\bs \xi\) states propagated to, then the propagator for an associated Hamiltonian can be truncated at order \(\hbar^0\) like this (up to a global phase). This can be determined by finding the orbit of the saddle point trajectories, or Grassmann equations of motion, which will generally be linear combinations of the Grassmann generators, \(\xi_p\), \(\xi_q\) and \(\xi_r\). We call this set of points the \emph{Weyl phase space points}, in analogy to the two-generator Wigner-Weyl-Moyal phase space, for reasons which will become clear.

The more relevant question becomes how many Weyl phase space \(\bs \xi\) points do we have to keep track of when we truncate like this. In particular, we would like a finite set. Such a finite phase space can be found by reexpressing the propagation of a Weyl symbol \(g(\xi)\) in terms of a transformation of its \(\xi\) argument:
\begin{widetext}
\begin{eqnarray}
  g_t(\xi) &=& \int U^*(\xi_1;t) g(\xi_2 + \xi_1 - \xi) U(\xi_2;t) e^{\frac{i}{\hbar} \Delta_3(\xi,\xi_1,\xi_2)} \text d^3 \xi_1 \text d^3 \xi_2 \nonumber \\
           &=& | \det(\bs I + \bs{\mathcal E})| \int g(\xi_2+\xi_1-\xi) e^{\frac{2}{\hbar} \left[i \sum_{k,l,m} \epsilon_{klm}\frac{b_k}{b}\tan\left(\frac{1}{2} b t\right) {\xi_1}_l {\xi_1}_m -i \epsilon_{klm}\frac{b_k}{b}\tan\left(\frac{1}{2} b t\right) {\xi_2}_l {\xi_2}_m  - (\xi_1 \xi_2 + \xi_2 \xi + \xi \xi_1) \right] } \text d^3 \xi_1 \text d^3 \xi_2 \nonumber \\
           &=& | \det(\bs I + \bs{\mathcal E})| \int g(\xi'_1) e^{\frac{2}{\hbar} \left[i \sum_{k,l,m} \epsilon_{klm}\frac{b_k}{b}\tan\left(\frac{1}{2} b t\right) {\xi'_2}_l ({\xi'_1+\xi})_m + \xi'_2 (\xi'_1-\xi)  \right] } \text d^3 \xi'_1 \text d^3 \xi'_2\\
           &=& | \det(\bs I + \bs{\mathcal E})| \int g(\xi'_1) \delta\left(i \sum_{k,m} \epsilon_{klm}\frac{b_k}{b}\tan\left(\frac{1}{2} b t\right) ({\xi'_1+\xi})_m + I ({\xi'_1+\xi}) \right) \text d^3 \xi'_1 \nonumber \\
  g_t(\xi) &=& g(\bs{\mathcal E}_{U} \xi), \nonumber
\end{eqnarray}
\end{widetext}
where \(\bs{\mathcal E}_U\) is related to the Hessian of the underlying Hamiltonian of the unitary gate \(U\):
\begin{equation}
  \label{eq:unitarymatrixE}
  \bs{\mathcal E}_{U} = \exp \left( i H_{U} \frac{\cev \partial^2}{\partial \bs \xi^2} t \right),
\end{equation}
for \(H_U = -\frac{i}{2} \sum_{k,l,m} \epsilon_{klm} b_k \xi_l \xi_m\) in the treatment above.

If we consider the \(\{\xi_p, \xi_q, \xi_r\}\) to be the three-generator Weyl phase points, then unitary matrices that take these elements to themselves (as opposed to linear combinations) are the discrete analogs of continuous operations that preserve phase space area. We will see that these gates have underlying Hamiltonians that are quadratic with restricted coefficients and that the Clifford gates fall into this class.

Note that in general such a matrix \(\bs{\mathcal E}_U\) exists for qubit evolution under all unitary gates \(U\) since the Grassmann algebra enforces all underlying Hamiltonians to be maximally quadratic. However, only Hamiltonians with appropriate coefficients produce \(\bs{\mathcal E}\) matrices that take these Weyl phase space points to themselves. The latter evolutions are of interest to us because they restrict us to only dealing with a finite set of \(\bs \xi\) Weyl phase space points and so allow us to truncate to order \(\hbar^0\) with a finite set of terms. However, in general, most unitary gates \(U\) do not possess such appropriate coefficients and so such a truncation to order \(\hbar^0\) produces an infinite set of points to keep track of (corresponding to all the quantum states they reach---i.e. their orbit in Hilbert space).

In other words, we define a \(3n\times 3n\) matrix \(\bs{\mathcal E}_{\hat O}\) such that
\begin{equation}
  \left(\begin{array}{c} \xi'_p\\ \xi'_r\\ \xi'_q \end{array}\right) = \bs{\mathcal E}_{\hat O} \left( \begin{array}{c} \xi_p\\ \xi_r\\ \xi_q \end{array}\right),
\end{equation}
for operators \(\hat O\) which can be treated to order \(\hbar^0\), where \(\bs{\mathcal E}_{\hat O}\) is related to the Hessian of \(H_{\hat O}\), the Hamiltonian associated with unitary gate \(\hat O\), which takes discrete Weyl phase space points \(\xi_\alpha\) to other Weyl phase space points \(\xi_\beta\) (and not linear combinations of them)~\cite{Kocia16}.

This is equivalent to the \(\hbar^0\) limit of the two-generator propagator which takes Weyl phase space points to themselves via a symplectic (area-preserving) matrix \(\bs{\mathcal M}\) and vector \(\bs \alpha\)~\cite{Almeida98}.

\subsection{Clifford Gates}

Again, consider the quadratic and real Hamiltonian,
\begin{equation}
  H(\xi) = -\frac{i}{2} \sum_{k,l,m} \epsilon_{klm} b_k \xi_l \xi_m.
  \label{eq:harmonicHam}
\end{equation}
As we found above, the corresponding propagator is
\begin{equation}
  \label{eq:qubitprop}
  \hat G(t) = \exp(-i t \hat H) = \hat I \cos \frac{b}{2} t - i \hat {\bs \sigma} \cdot \bs n \sin \frac{b}{2} t,
\end{equation}
where \(b = |\bs b|\) and \(\bs n = \bs b/b\).

From Eq.~\ref{eq:eomGrassmann}, the equations of motion of the Grassmann elements are
\begin{equation}
  \frac{\text d}{\text d t} \xi_k = i H \frac{\cev \partial}{\partial \xi_k} = \epsilon_{k l m} b_l \xi_m.
\end{equation}

The Clifford gate set consists of the one-qubit Hadamard and phase shift gates and the two-qubit controlled-not (CNOT) gate.

For the one-qubit Hadamard \(\hat F\) gate,
\begin{equation}
  \hat F = \frac{1}{\sqrt{2}} \left( \begin{array}{cc}1& 1\\ 1& -1 \end{array} \right) = \frac{1}{\sqrt{2}} ( \hat Z + \hat X ),
\end{equation}
it follows from Eq.~\ref{eq:qubitprop} that if we set \(t=\pi\), then \(b = 1\) and \(\bs n = (1, 0, 1)/\sqrt{2}\). Hence,
\begin{equation}
  H_{\hat F} = -\frac{i}{\sqrt{2}} ( \xi_r \xi_q + \xi_p \xi_r).
\end{equation}

Hence, under the Hadamard the Grassmann elements evolve in time under the equations of motion
\begin{equation}
  \frac{\text d}{\text d t}(\xi_p, \xi_q, \xi_r) = \frac{1}{\sqrt{2}} (\xi_r, -\xi_r, -\xi_p + \xi_q),
\end{equation}
which, when solved for the time \(t=\pi\), or using Eq.~\ref{eq:unitarymatrixE}, produce:
\begin{equation}
  (\xi'_p, \xi'_q, \xi'_r) = (\xi_q, \xi_p,-\xi_r ).
\end{equation}

For the one-qubit phase shift gate \(\hat P\),
\begin{equation}
  \hat P = \left(\begin{array}{cc}1 & 0\\ 0 & i \end{array} \right) = \frac{e^{\frac{\pi i}{4}}}{\sqrt{2}} (\hat I - i \hat Z),
\end{equation}
it follows that, ignoring its overall phase, setting \(t=\pi/2\), implies that \(b = 1\) and \(\bs n = (0, 0, 1)\). Hence,
\begin{equation}
  H_{\hat P} = -i \xi_p \xi_r.
\end{equation}
The associated equations of motion are:
\begin{equation}
  \frac{\text d}{\text d t} (\xi_p,\xi_q,\xi_r) = - (\xi_r,0,-\xi_p).
\end{equation}

Now solving for the equations of motion for the time \(t=\pi/2\), or using Eq.~\ref{eq:unitarymatrixE}, reveals that
\begin{equation}
  (\xi'_p,\xi'_q,\xi'_r) = (-\xi_r, \xi_q, \xi_p).
\end{equation}

For the two-qubit CNOT gate \(\hat C_{ab}\),
\begin{eqnarray}
  \hat C_{ab} &=& \left(\begin{array}{cccc} 1& 0& 0& 0\\ 0& 1& 0& 0\\ 0& 0& 0& 1\\ 0& 0& 1& 0 \end{array}\right)\\
  &=& \frac{1}{2}(\hat I + \hat X_b + \hat Z_a - \hat X_b \hat Z_a).
\end{eqnarray}
    
It follows that since \(\Hat C_{ab}^2 = \hat I\),
\begin{equation}
  e^{i \theta \hat C_{ab}} = \hat I \cos \theta + i \hat C_{ab} \sin \theta.
\end{equation}
Setting \(\theta = \pi / 2\) reveals that
\begin{equation}
  \hat C_{ab} = -i \exp i \frac{\pi}{4} \left( \hat I + \hat Z_b + \hat X_a - \hat Z_b \hat X_a \right),
\end{equation}
and so, setting \(t=\pi\)
\begin{equation}
  H_{\hat C_{ab}} = -\frac{i}{4} \left( \xi_{p_b}\xi_{r_b} + \xi_{r_a} \xi_{q_a} + i\xi_{p_b}\xi_{r_b} \xi_{r_a}\xi_{q_a} \right)
\end{equation}
(up to a constant). Rewriting this as \(\frac{1}{4}\left(1 - i \xi_{p_b} \xi_{r_b} \right) \left( 1 - i \xi_{r_a} \xi_{q_a} \right)\) we find that this can be thought of as a product of Eq.~\ref{eq:harmonicHam}, \(H_a(\xi_a) H_b(\xi_b)\), where \(b_a = b_b = 1\), \(\bs n_a = (1,0,0)\), and \(\bs n_b = (0,0,1)\).

The associated equations of motion that are non-zero are:
\begin{eqnarray}
  \der{}{t} \xi_{r_a} &=& -\frac{1}{4} \xi_{q_a} (1+i \xi_{p_b} \xi_{r_b}), \nonumber\\
  \der{}{t} \xi_{q_a} &=& \frac{1}{4} \xi_{r_a} (1+i \xi_{p_b} \xi_{r_b}), \nonumber\\
  \der{}{t} \xi_{p_b} &=& -\frac{1}{4} \xi_{r_b}(1+i \xi_{r_a} \xi_{q_a}), \nonumber\\
  \label{eq:cnotevolution}
  \der{}{t} \xi_{r_b} &=& \frac{1}{4} \xi_{p_b}(1+i \xi_{r_a} \xi_{q_a}), \\
  \der{}{t} \left(\xi_{r_a}\xi_{p_b}\xi_{r_b}\right) &=& -\frac{i}{4} \xi_{q_a}, \nonumber\\
  \der{}{t} \left(\xi_{q_a}\xi_{p_b}\xi_{r_b}\right) &=& \frac{i}{4} \xi_{r_a}, \nonumber\\
  \der{}{t} \left(\xi_{p_b}\xi_{p_a}\xi_{r_a}\right) &=& -\frac{i}{4} \xi_{r_b}. \nonumber\\
  \der{}{t} \left(\xi_{r_b}\xi_{p_a}\xi_{r_a}\right) &=& \frac{i}{4} \xi_{p_b}, \nonumber
\end{eqnarray}

Solving these equations of motion for the time \(t=\pi\) reveals that the only odd monomials that change are:
\begin{eqnarray}
  \xi_{r_a} &\rightarrow& i \xi_{r_a} \xi_{r_b} \xi_{p_b}, \nonumber\\
  \xi_{r_a}\xi_{p_b} \xi_{r_b} &\rightarrow& -i \xi_{r_a}, \nonumber\\
  \xi_{q_a} &\rightarrow& i \xi_{q_a} \xi_{r_b} \xi_{p_b}, \nonumber\\
  \label{eq:cnotGrassmannevolution}
  \xi_{q_a}\xi_{p_b} \xi_{r_b} &\rightarrow& -i \xi_{q_a},\\
  \xi_{r_b} &\rightarrow& i \xi_{r_b} \xi_{q_a} \xi_{r_a}, \nonumber\\
  \xi_{r_b}\xi_{q_a} \xi_{r_a} &\rightarrow& -i \xi_{r_b}, \nonumber\\
  \xi_{p_b} &\rightarrow& i \xi_{p_b} \xi_{q_a} \xi_{r_a}, \nonumber\\
  \xi_{p_b}\xi_{q_a} \xi_{r_a} &\rightarrow& -i \xi_{p_b}. \nonumber
\end{eqnarray}
As we will see in Section~\ref{sec:stabstates}, the factor \(\frac{1}{2}(1-i \xi_{p_b} \xi_{r_b})\) is the projector on the \(+1\) eigenstate of the \(\hat Z_b\) operator while the factor \(\frac{1}{2}(1-i \xi_{r_q} \xi_{q_a})\) is the projector on the \(+1\) eigenstate of the \(\hat X_a\) operator. In other words, Eq.~\ref{eq:cnotevolution} shows that the evolution of qubit \(a\) is conditioned on qubit \(b\) being in a position state (\(\frac{1}{2}\left(1 - i \xi_{p_b} \xi_{r_b} \right)\)), and the evolution of qubit \(b\) is conditioned on qubit \(a\) being in a momentum state (\(\frac{1}{2}\left(1 - i \xi_{p_a} \xi_{r_a} \right)\)). A similar structure was observed for odd \(d\)

As can be seen from the equations of motion for the three Clifford gates, their evolution takes the discrete states \(\xi_k\) to other discrete states \(\xi'_k\). This is due to the Hamiltonians associated with these operators being quadratic with appropriate coefficients (for \(\hbar = 2\) and \(t\) equal to fractions of \(\pi\)). As a result, following the argument made in Section~\ref{sec:vanvleckprefactor}, it follows that they can be treated by the Weyl propagator truncated at order \(\hbar^0\). We will shortly see that this is generally not the case for a gate set that produces universal quantum computation.

\subsection{T-Gate}
To extend the Clifford gate set to a universal quantum gate set, it is only necessary to add the T-gate, defined as the square root of the phase-shift gate:

\begin{equation}
  \hat T = \left(\begin{array}{cc}1 & 0\\ 0 & e^{\frac{\pi i}{4}} \end{array} \right) = \frac{1}{2}\left( (1 + e^{-\frac{\pi i}{4}}) \hat I + (1 - e^{-\frac{\pi i}{4}}) \hat Z \right).
\end{equation}
For \(t=\pi/2\), it follows that the Hamiltonian is the same as for the phase-shift, except halved:
\begin{equation}
  H_{\hat T} = -\frac{i}{2} \xi_p \xi_r.
\end{equation}
The equations of motion are:
\begin{equation}
  \frac{\text d}{\text d t} (\xi_p\xi_q,\xi_r) = -\frac{1}{2} (\xi_r,0,-\xi_p).
\end{equation}

Now solving for the equations of motion for the unit time interval reveals that that 
\begin{eqnarray}
  \xi'_p &=& \frac{1}{\sqrt{2}}(\xi_p - \xi_r),\\
  \xi'_r &=& \frac{1}{\sqrt{2}}(\xi_p + \xi_r),
\end{eqnarray}
and
\begin{equation}
  \xi'_q = \xi_q.
\end{equation}

Unlike the Clifford operators, the T-gate takes the discrete states \(\xi_k\) to linear combinations of \(\xi_k\). This is due to the fact that its corresponding Hamiltonian is no longer quadratic with coefficients that are integers. Thus, in terms of the stabilizer operator basis, it cannot be treated by the Weyl propagator truncated at order \(\hbar^0\), and the full \(\hbar^1\) expression must be used; the T-gate takes the Weyl phase space points to linear combinations of Weyl phase space points.

\subsection{Phase Space and Non-Negative Probability Distributions}
\label{sec:phasespace}

In the two-generator Wigner-Weyl-Moyal formalism for odd \(d\), the Wigner function of a stabilizer state can be interpreted as the real non-negative coefficients in front of \(\hat R(\bs x)\), indexed by \(\bs x\), in the expansion of the density operator basis given by the \(\hat R\)'s:
\begin{eqnarray}
  \label{eq:twogenlincombo}
  \hat g &=& \sum_{\bs x} g(\bs x) \hat R(\bs x)\\
         &=& g(0,0) \hat R(0,0) + g(1,0) \hat R(1,0) + \ldots.\nonumber\\
\end{eqnarray}
The Clifford gates are \emph{covariant} in terms of these operators, i.e. they take stabilizer density states' (non-negative) coefficients in front of these operators and permute them. This permutation can be captured by a symplectic matrix \(\bs{\mathcal{M}}\). In the continuous world a symplectic matrix can be described as ``area-preserving'', while in the discrete world it is perhaps more appropriate to describe its action as a permutation of coefficients, which by defintion is bijective and hence lossless.

In the three generator Wigner-Weyl-Moyal formalism, a very similar interpretation is possible. However, since qubit Clifford gates are also a three-design, it is not possible to express their action on any such operator basis in a covariant manner~\cite{Zhu15}. Nevertheless, a natural analog to Eq.~\ref{eq:twogenlincombo} is:
\begin{eqnarray}
  \label{eq:threegenlincombo}
  \hat g &=& g(\hat \xi)\\
         &=& 1 + g_{pr} \hat \xi_p \hat \xi_r + g_{rp} \hat \xi_r \hat \xi_p \nonumber\\
         && + g_{pq} \hat \xi_p \hat \xi_q + g_{qp} \hat \xi_q \hat \xi_p \nonumber\\
         && + g_{rq} \hat \xi_r \hat \xi_q + g_{qr} \hat \xi_q \hat \xi_r. \nonumber
\end{eqnarray}

Simplifying the products of Pauli matrices above, we can write
\begin{eqnarray}
  \label{eq:threegenlincombo2}
  \hat g &=& g(\hat \xi)\\
         &=& 1 + g_p \hat \xi_p + g_r \hat \xi_r + g_q \hat \xi_q \nonumber\\
         && - g_{\text{-}p} \hat \xi_p - g_{\text{-}r} \hat \xi_r - g_{\text{-}q} \hat \xi_q.
\end{eqnarray}

In this way, we can express the operator basis as:
\begin{eqnarray}
  \{1,\, \pm \hat \xi_p,\, \pm \hat \xi_r,\, \pm \hat \xi_q\},
\end{eqnarray}
for one degree of freedom (one qubit). Note that plus and minus signs are only necessary when no longer using the even representation of this operator basis, which is perhaps less cosmetically appealing. More degrees of freedom correspond to products of these phase space points. For instance, for two qubits they are:
\begin{eqnarray}
  &&\{\hat I_a \hat I_b,\, \pm \hat \xi_{p_a} \hat I_b,\, \pm \hat \xi_{r_a} \hat I_b,\,  \pm \hat \xi_{q_a} \hat I_b, \pm \hat I_a \hat \xi_{p_b},\,  \pm \hat I_a \hat \xi_{r_b},\, \pm \hat I_a \hat \xi_{q_b}, \nonumber\\
  &&\quad \pm \hat \xi_{p_a} \hat \xi_{p_b},\, \pm \hat \xi_{p_a} \hat \xi_{r_b},\, \pm \hat \xi_{p_a} \hat \xi_{q_b}, \pm \hat \xi_{r_a} \hat \xi_{p_b},\, \pm \hat \xi_{r_a} \hat \xi_{r_b},\, \pm \hat \xi_{r_a} \hat \xi_{q_b}, \nonumber\\
  && \quad \pm \hat \xi_{q_a} \hat \xi_{p_b},\, \pm \hat \xi_{q_a} \hat \xi_{r_b},\, \pm \hat \xi_{q_a} \hat \xi_{q_b}\}.
\end{eqnarray}
These are nothing more than the well-known stabilizer operators~\cite{Gottesman98}! We thus see that in the two-generator Wigner-Weyl-Moyal formalism the operator basis consists of the reflection \(\hat R\)-basis operators whereas in the three-generator Wigner-Weyl-Moyal formalism the operator basis consists of stabilizer operators. While the \(\hat R\)-operators are Hilbert-Schmidt orthogonal, the stabilizer operators are not. This is expected since we know a Hilbert-Schmidt orthogonal operator basis cannot remain invariant under Clifford operators as they are a three-design~\cite{Zhu15}.

As a result, it is easy to construct a local hidden variable theory to describe qubit stabilizer propagation using a non-negative probability distribution related to the three-generator Wigner-Weyl-Moyal formalism. We define such a probability distribution \(\bar g\), where \(\bar g:\mathbb Z/6\mathbb Z \rightarrow \mathbb R\), to consist of the coefficients in front of the stabilizer operators of \(g(\hat \xi)\) (excluding the trivial identity operator):
\begin{equation}
  \label{eq:probdistrogbar}
  \bar g = \left( g_p,\, g_r,\, g_q,\, g_{\text{-}p},\, g_{\text{-}r},\, g_{\text{-}q}\right)
\end{equation}
for one degree of freedom. This takes integers to the real line and is non-negative for stabilizer states. In particular, suitably normalized, \(\sum_x \bar g(x) = 1\) and can be interpreted as a classical probability on the phase space of even Grassmann elements. Wallman \emph{et al}. found the same non-negative probability distribution for stabilizer states by considering the octahedral group for one qubit~\cite{Wallman12}.

For a one-qubit Clifford gate \(\hat V \in \{\hat F, \hat P\}\), it follows that there exists a \(6 \times 6\) permutation matrix \(\bs{\mathcal P}_{\hat V}\) such that
\begin{equation}
  \label{eq:threegensymplecticprop}
  \bar g(\bs x) \underset{\hat V}{\rightarrow} \bar g(\bs{\mathcal P}_{\hat V} \bs x).
\end{equation}

In particular,
\begin{equation}
  \bs{\mathcal P}_{\hat F} = \left(
    \begin{array}{cccccc}
      0&0&1&0&0&0\\
      0&0&0&0&1&0\\
      1&0&0&0&0&0\\
      0&0&0&0&0&1\\
      0&1&0&0&0&0\\
      0&0&0&1&0&0
    \end{array}
\right),
\end{equation}
and
\begin{equation}
  \bs{\mathcal P}_{\hat P} = \left(
  \begin{array}{cccccc}
    0&0&0&0&1&0\\
    1&0&0&0&0&0\\
    0&0&1&0&0&0\\
    0&1&0&0&0&0\\
    0&0&0&1&0&0\\
    0&0&0&0&0&1
  \end{array} \right).
\end{equation}

Similarly, for two degrees of freedom (two qubits):
\begin{eqnarray}
  \label{eq:2dprobdistrogbar}
  \bar g &=& (g_{p0}, g_{pp}, g_{pr}, g_{pq}, g_{r0}, g_{rp}, g_{rr}, g_{rq}, \nonumber\\
         && g_{q0}, g_{qp}, g_{qr}, g_{qq}, g_{\text{-}p0}, g_{\text{-}pp}, g_{\text{-}pr}, g_{\text{-}pq},\\
         &&g_{\text{-}r0}, g_{\text{-}rp}, g_{\text{-}rr}, g_{\text{-}rq}, g_{\text{-}q0}, g_{\text{-}qp}, g_{\text{-}qr}, g_{\text{-}qq}). \nonumber
\end{eqnarray}
For the two-qubit Clifford gate \(\hat C_{ab}\), it follows that there exists a \(30\times 30\) permutation matrix \(\bs{\mathcal P}_{\hat C_{ab}}\) that obeys Eq.~\ref{eq:threegensymplecticprop}.

The Clifford operators in the previous section take the Weyl phase space points in Eq.~\ref{eq:probdistrogbar} and Eq.~\ref{eq:2dprobdistrogbar} to themselves. In other words, they conserve phase space area, and, if framed in terms of two conjugate degrees of freedom, would be a symplectic transformation.

Lastly, as in the two-generator formalism, the indices that enumerate the operator basis can be associated with Weyl phase space points. It follows here that the phase space grows as \(2 \times 4^n-1\) for \(n\) qubits. We will show in the next section that stabilizer states are only defined, i.e. have positive probabilities, on \(2 \times 3^n\) of these points. Note that this does not mean that there are \({2 \times 4^n -1}\choose{2 \times 3^n}\) stabilizer states possible for \(n\) qubits, as most combinations are not allowed. Indeed, the relationship between the number of possible stabilizer states with \(n\) is \(2^{\left(\frac{1}{2}+\mathcal O(1)\right)n^2}\)~\cite{Aaronson04}. 

\subsection{Stabilizer State Representation}
\label{sec:stabstates}

The stabilizer states for one qubit correspond to the six eigenstates of the \(\hat X\), \(\hat Y\), and \(\hat Z\) operators. As such their Weyl symbols, or Wigner functions, are easy to find and correspond to
\begin{eqnarray}
  \xi_{p_\pm} &\equiv& \frac{1}{2} (1 \pm i \xi_r \xi_q),\\
  \xi_{q_\pm} &\equiv& \frac{1}{2} (1 \pm i \xi_p \xi_r),\\
  \xi_{r_\pm} &\equiv& \frac{1}{2} (1 \pm i \xi_p \xi_q).
\end{eqnarray}
The operators corresponding to these Wigner functions can be found by replacing the Grassmann elements by their corresponding scaled Pauli matrices and \(1\) with the identity matrix. The resultant operators are the usual projectors onto the associated eigenvectors.

To extend this definition of stabilizer states to more than one qubit, we use the following definition of stabilizer states:
\begin{definition}
  A stabilizer state is defined as any state reached by Clifford gates from an initially prepared \(\ket{0\ldots 0}\) state (in the \(Z\) basis).
\end{definition}
Since we have shown that Clifford gates \(\hat O\) are captured at order \(\hbar^0\), or equivalently, take positive elements---the stabilizer states---to themselves, the \(\ket{0\ldots 0}\) state is a tensor product of \(\frac{1}{2}(1+\xi_p \xi_r)\) non-negative states, it follows that \(\hat O \ket{0\ldots 0}\) is non-negative as well.

\section{Wigner Function with Two Generators}
\label{sec:twogenerators}

As discussed in the Introduction, the usual two-generator discrete Weyl symbol is formulated as a periodization and discretization of the continuous Weyl formalism. However, if the same Weyl-Heisenberg operators (Eq.~\ref{eq:weylheisenberg}) that are used to define the translation operators in odd \(d\) are applied to even \(d\), they no longer form a subgroup of \(SU(d)\). As a result, we will show that their their dual \(\hat R_{ij}\) operators (see Eq.~\ref{eq:reflectionopforoddd}), can no longer be interpreted as reflection operators in phase space that take each Weyl phase space point \(\bs \xi\) to another, independently of all others.

The generalized phase space translation operator (often called the Weyl operator) for qudits with prime or odd \(d\) can be defined as a product of the shift and boost:
\begin{equation}
\hat {T}(\bs \lambda_p, \bs \lambda_q) = \omega^{-\bs \lambda_p \cdot \bs \lambda_q (d+1)/2} \hat {Z}^{ \bs \lambda_p} \hat {X}^{ \bs \lambda_q},
\end{equation}
where \(\bs \lambda \equiv (\bs \lambda_p, \bs \lambda_q) \in \mathbb{Z}/d\mathbb{Z}\) define the chord phase space, and \(\omega \equiv \exp 2 \pi i/ d\). Notice that for \(d=2\) and setting \(\omega = \exp \pi i/2\), \(\hat T(1,1) = \hat Y\), and so the \(\hat T\) operators correspond to the Pauli matrices and identity matrix.

The translation operators obey the following group relations of the Weyl-Heisenberg group:
\begin{equation}
\hat {T}\left(\bs \lambda_2\right) \hat {T}\left(\bs \lambda_1\right) = \omega^{\bs \lambda_1^T \bs{\mathcal J} \bs \lambda_2} \hat {T}\left(\bs \lambda_1 + \bs \lambda_2\right),
\end{equation}
for
\begin{equation}
  \bs{\mathcal J} = \left(\begin{array}{cc} 0 & -\mathbb{I}_n\\ \mathbb{I}_n & 0\end{array}\right),
\end{equation}
and \(\mathbb I_n\) the \(n\times n\) identity matrix.

We define \(\hat R(x)\) as the symplectic Fourier transform of \(\hat T(\lambda)\):

\begin{equation}
  \label{eq:reflectionopforoddd}
\hat {R}(\bs x_p, \bs x_q) = d^{-n} \sum_{\substack{\bs \lambda_p, \bs \lambda_q \in \\ (\mathbb{Z} / d \mathbb{Z})^{ n}}} e^{\frac{2 \pi i}{d} (\bs \lambda_p, \bs \lambda_q) \bs{\mathcal J} (\bs x_p, \bs x_q)^T} \hat {T}(\bs \lambda_p, \bs \lambda_q).
\end{equation}

For odd \(d\) these \(\hat R(\bs x)\) operators can be seen to be a reflection (actually an inversion) around \(\bs x\). Their trace with an operator \(\hat O\) defines the two-generator Weyl symbol of \(\hat O\):
\begin{equation}
  O_x(\bs x) = Tr(\hat R(\bs x)^\dagger \hat O).
\end{equation}
When this is expressible for a unitary gate as a single exponential with a quadratic argument (the action) with integer coefficients, the operator can be treated at order \(\hbar^0\)~\cite{Kocia16}.

As a result, for the operators \(\hat O\) that can be treated at order \(\hbar^0\), we can define the \(2n\)-vector \(\mathcal {\bs \alpha}_{\hat O}\) and an \(2n\times 2n\) symplectic matrix \(\bs{\mathcal M}_{\hat O}\) with entries in \(\mathbb Z/d \mathbb Z\) such that~\cite{Rivas99}:
\begin{equation}
  \left( \begin{array}{c}\bs x_{p'}\\ \bs x_{q'}\end{array} \right) = {\bs{\mathcal M}}_{\hat O} \left[ \left( \begin{array}{c}\bs x_p\\ \bs x_q \end{array}\right) + \frac{1}{2} {\bs \alpha}_{\hat O} \right] + \frac{1}{2} {\bs \alpha}_{\hat O},
  \label{eq:quadmap}
\end{equation}
which is associated with the quadratic action
\begin{equation}
  \label{eq:quadcentgenfunction}
  S_{\hat O}(\bs x_p, \bs x_q) =  \bs \alpha_{\hat O}^T \bs {\mathcal J} \left(\begin{array}{c}\bs x_p\\ \bs x_q\end{array}\right) + (\bs x_p, \bs x_q) \bs{\mathcal B}_{\hat O} \left(\begin{array}{c}\bs x_p\\ \bs x_q\end{array}\right),
\end{equation}
where \(\bs{\mathcal B}_{\hat O}\) is a real symmetric \(2n\times 2n\) matrix that is related to \(\bs{\mathcal M}_{\hat O}\) by the Cayley parameterization~\cite{Golub12}:
\begin{eqnarray}
  \label{eq:symplecticcayleyparam}
  \bs {\mathcal J} \bs{\mathcal B}_{\hat O} &=& \left( 1 + \bs {\mathcal M}_{\hat O} \right)^{-1} \left( 1 - \bs {\mathcal M}_{\hat O} \right) \\
  &=& \left( 1 - \bs {\mathcal M}_{\hat O} \right)  \left( 1 + \bs {\mathcal M}_{\hat O} \right)^{-1}. \nonumber
\end{eqnarray}

This is no longer the case for \(d=2\). Operators \(\hat O\) that can be treated at order \(\hbar^0\) cannot be described by a simple matrix \(\mathcal{\bs M}\) and vector \(\alpha\). This is because \(\hat T\) and \(\hat R\) don't accomplish the expected translation and reflections. The \(\hat R\) operator, upon which the center (Wigner) representation is built, from Eq.~\ref{eq:reflectionopforoddd} is:
\begin{eqnarray}
  &&\hat R(x_p, x_q) \\
  &=&\frac{1}{2}\left[(-1)^{x_q}\hat Z + (-1)^{x_p} \hat X + i (-1)^{x_p + x_q}\hat X \hat Z + \hat I\right]. \nonumber
\end{eqnarray}
This agrees with Eq.~\(10\) in Wootters' original derivation~\cite{Wootters87}. When applied to the stabilizer state \(\ket{0}\) (in the \(Z\) basis) for instance, we find:
\begin{eqnarray}
  &&\hat R(\bs x) \ket{0}\\
  &=& \frac{1}{2} \left[ \left((-1)^{x_q}+1\right)\ket{0} + \left((-1)^{x_p} - i (-1)^{x_p + x_q}\right)\ket{1}\right].\nonumber
\end{eqnarray}
It follows that
\begin{equation}
  \hat R(\bs x) \ket{0} = \begin{cases} \ket{0} + \frac{1-i}{2} \ket{1} & \text{for}\quad \bs x =(0,0),\\ \frac{1+i}{2} \ket{1} & \text{for}\quad \bs x =(0,1),\\ \ket{0} - \frac{1-i}{2} \ket{1} & \text{for}\quad \bs x =(1,0), \, \text{and}\\ \frac{1-i}{2} \ket{1} & \text{for}\quad \bs x =(1,1).\end{cases}
\end{equation}
Therefore, this only takes a stabilizer state to another for \(\bs x = (1,1)\) and \(\bs x = (0,1)\).

As a result, the two-generator Wigner-Weyl-Moyal formalism, which establishes a relationship between the center of reflection operators and ``center'' representations (or Weyl symbols), is not possible for \(d=2\). Nevertheless, these discrete \(d=2\) Wigner functions are a perfectly valid representation of a quantum state, they just no longer have the usual Wigner-Weyl-Moyal (center and chord) formalism underpinning them. Furthermore, when expressed as symplectic matrices, the Clifford operators in this discrete Wigner formalism are not state-independent, as we shall show shortly.

To analyze the effect of the Clifford group gates on stabilizer states in this two-generator representation, we will instead first consider their three-generator Grassmann representation and use the following map between the Grassmann three-generator algebra \(\mathcal G_3\) stabilizer states and those of the two-generator algebra \(\mathcal C_2\):
\begin{eqnarray}
  1-i\xi_r \xi_q \mapsto \delta_{p,1}\\
  1+i\xi_r \xi_q \mapsto \delta_{p,0}\\
  1-i\xi_p \xi_r \mapsto \delta_{q,1}\\
  1+i\xi_p \xi_r \mapsto \delta_{q,0}\\
  1-i\xi_p \xi_q \mapsto \delta_{p,q}\\
  1+i\xi_p \xi_q \mapsto \delta_{p,1\oplus q},
\end{eqnarray}
where \(\oplus\) denotes mod \(2\) arithmetic and \(p,\,q\in\{0,1\}\).
This mapping is illustrated in Fig.~\ref{fig:qubitstabstates}.

\subsection{Stabilizer States}

\begin{figure}[ht]
  \includegraphics[scale=1.0]{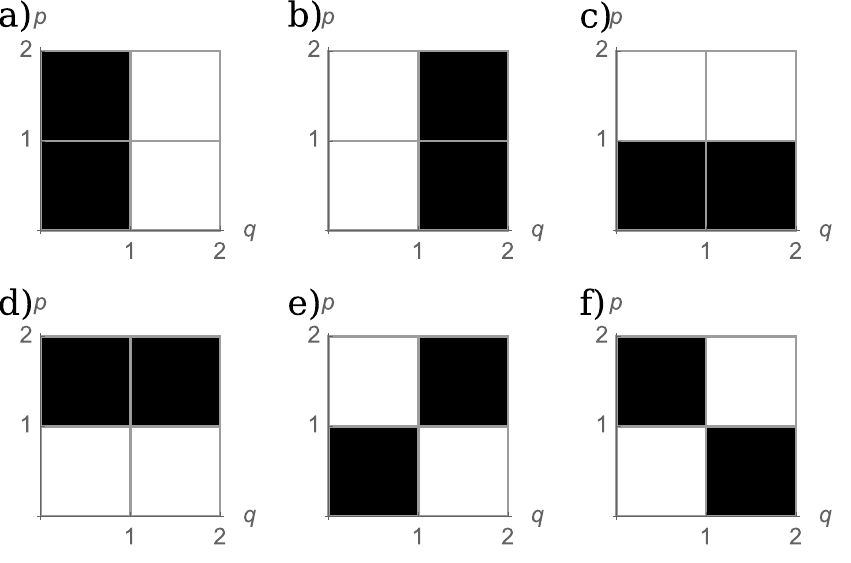}
  \caption{Qubit stabilizer states. a) and b) are position or \(\hat Z\)-states with Weyl symbol \(1\pm i \xi_p\xi_r\), c) and d) are momentum or \(\hat X\)-states with Weyl symbol \(1\pm i \xi_q \xi_r\), and e) and f) are diagonal or \(\hat Y\)-states with Weyl symbol \(1\pm i \xi_p \xi_q\).}
  \label{fig:qubitstabstates}
\end{figure}

The stabilizer states under the two-generator formalism are non-negative (see Fig.~\ref{fig:qubitstabstates}). Indeed, we can apply Corollary~\(3\) from~\cite{Kocia16} to the prime \(d=2\) case here. This Corollary shows that a ``mixed'' representation is always possible: where each degree of freedom is expressed in either the \(p\)- or \(q\)-basis:
\begin{corollary}
  \label{cor:stabstatemixedrep}
For prime \(d\), if \(\Psi\) is a stabilizer state for \(n\) qudits, then there always exists a mixed representation in position and momentum such that:
\begin{equation}
  \Psi_{\theta_{\beta \bs x},\eta_{\beta \bs x}}(\bs x) = \frac{1}{\sqrt{d}} \exp \left[\frac{2 \pi i}{d} \left( \bs x^T \bs \theta_{\beta \bs x} \bs x + \bs \eta_{\beta \bs x} \cdot \bs x \right) \right],
  \label{eq:stabstatemixedrep}
\end{equation}
where \(x_i\) can be either \(p_i\) or \(q_i\).
\end{corollary}
The proof for this follows the same lines as those in~\cite{Kocia16}.

Thus there are six one-qubit stabilizer states: the two position or \(\hat Z\)-states, the two momentum or \(\hat X\)-states (\(\frac{1}{\sqrt{2}} (\ket 0 \pm \ket 1\)), and the two diagonal or \(\hat Y\)-states (\(\frac{1}{\sqrt{2}} (\ket 0 \pm i \ket 1\)) shown in Fig.~\ref{fig:qubitstabstates}. As a result of their exponential form with imaginary argument in Eq.~\ref{eq:stabstatemixedrep}, they are non-negative.

Though the stabilizer states are non-negative in this two-generator case, the Clifford gates are not single exponentials with quadratic arguments that can be treated at order \(\hbar^0\). In particular, the phase shift gate in center representation is
\begin{equation}
  \label{eq:twogenphaseshiftcenterrep}
  P_x(x_p, x_q) = \frac{1}{\sqrt{2}}\left(e^{\frac{\pi i}{2}} + e^{\pi i x_q}\right),
\end{equation}
up to an overall phase, while the Hadamard gate in center representation is
\begin{equation}
  \label{eq:twogenhadamardcenterrep}
  F_x(x_p, x_q) = \frac{1}{\sqrt{2}}\left(e^{-\pi i x_p} + e^{\pi i x_q}\right).
\end{equation}
The sum over more than one exponential term is emblematic of the fact that they cannot be rewritten in terms of a single symplectic \(\bs{\mathcal M}\) matrix and vector \(\bs \alpha\) acting on \((x_p, x_q)\)~\cite{Kocia16}.

We recall the evolution under the Hadamard gate for the three generators
\begin{equation}
  \label{eq:GrassmannHadmardevolution}
  \left(\begin{array}{c}\xi'_p\\ \xi'_q\\ \xi'_r\end{array}\right) = \left(\begin{array}{ccc} 0& 1& 0\\ 1& 0 & 0\\ 0& -1& 0 \end{array} \right) \left(\begin{array}{c}\xi_p\\\xi_q\\ \xi_r\end{array}\right).
\end{equation}
This simply describes an exchange of \(p\)- and \(q\)-states, and as such can be described by an evolution on the two generators \(\hat p\) and \(\hat q\) by using the following stability matrix
\begin{equation}
  \label{eq:stabmathad}
\bs{\mathcal M}_{\hat {F}} = \left( \begin{array}{cc} 0 & 1\\ -1 & 0 \end{array} \right),
\end{equation}
when the stabilizer states being propagated are \(p\)- or \(q\)-states. However, for \(r\)-states, Eq.~\ref{eq:GrassmannHadmardevolution} shows that the state must instead be translated by \((1,0)\) or \((0,1)\), since the Hadamard operator exchanges the \(r\)-states themselves, and the above \(\mathcal M_F\) matrix leaves the \(r\)-states invariant on \(p\)-\(q\) Weyl phase space.

Similarly, under the phase shift gate the three generators evolve by:
\begin{equation}
  \left(\begin{array}{c}\xi'_p\\ \xi'_q\\ \xi'_r\end{array}\right) = \left(\begin{array}{ccc} 0& 0& -1\\ 0& 1& 0\\ 1& 0& 0\end{array} \right) \left(\begin{array}{c}\xi_p\\ \xi_q\\ \xi_r\end{array}\right).
\end{equation}
Therefore, the phase shift is a \(q\)-shear in two-generator space, which leaves the \(q\)-states alone and shears the \(p\)-state so they become \(r\)-states.  This can be expressed in the two-generator picture by the following stability matrix when the states being propagated are \(p\)- or \(q\)-states:
\begin{equation}
\bs{\mathcal M}_{\hat {P}} = \left( \begin{array}{cc} 1 & 1\\ 0 & 1 \end{array} \right),
\end{equation}
Again, for \(r\)-states, this evolution is incorrect. For \(r\)-states the phase-shift gate acts as a \(p\)-shear in two-generator Weyl space, which takes \(r\)-states to \(p\)-states. A \(p\)-shear is equivalent to a \(q\)-shear followed by a translation by \((1,0)\) or \((0,1)\) (when the boundary conditions are periodic). Therefore, \(r\)-state evolution must be followed by such a translation in the two-generator Weyl picture if \(\mathcal M_{\hat P}\) is used.

To summarize, for the one-qubit Clifford gates (consisting of the Hadamard and phase shift gates), when the state is a position or momentum state the corresponding stability matrix is applied by itself,
\begin{equation}
  W_{\hat O \ket q, \hat O \ket p}( \bs x) = W_{\ket q, \ket p}(\bs{\mathcal M}_{\hat O} \bs x),
\end{equation}
while for the state \(\ket r = \frac{1}{\sqrt{2}} (\ket 0 \pm i \ket 1)\), the translation vector \(\bs r\) is also applied:
\begin{equation}
  W_{\hat O \ket r}( \bs x) = W_{\ket r}\left(\bs{\mathcal M}_{\hat O} \bs x + \bs r \right),
\end{equation}
where \(\bs r\) can be equivalently \((1, 0)\) or \((0,1)\). In the Aaronson-Gottesman tableau algorithm~\cite{Aaronson04}, this is equivalent to the binary arithmetic : ``set \(r_i := r_i \oplus x_{i_a} z_{i_a}\)'' since \(x_{i_a} = z_{i_a} = 1\) iff qubit \(a\) is in an \(r\)-state.

Since the two-generator center representations for the phase-shift gate in Eqs.~\ref{eq:twogenphaseshiftcenterrep} and~\ref{eq:twogenhadamardcenterrep} aren't in a single exponential term as for odd \(d\) (see ~\cite{Kocia16}), their path integral treatment requires terms up to \(\hbar^1\) in general. However, we have shown that it is still possible to propagate a stabilizer state classically with a single corresponding \(\bs{\mathcal M}\) matrix of the Clifford gate operator if this is done state-dependently, in agreement with Aaronson-Gottesman's tableau algorithm~\cite{Aaronson04}. More precisely, though an all encompassing matrix \(\mathcal M_{\hat O}\) that characterizes the Clifford gate \(\hat O\) evolution cannot be defined, we have shown that a state-dependent vector \(\bs r\) remedies the problem.

Similarly, recalling the evolution for the three generators under the controlled-not gate to be Eq.~\ref{eq:cnotGrassmannevolution}, we see that this is an inversion of \(q_a\)- and \(r_a\)-states if qubit \(b\) is in the \({q_+}_b\) state (the \(+1\) eigenstate of \(\hat Z_b\)) and an inversion of \(p_b\)- and \(r_b\)-states if qubit \(a\) is in the \({p_+}_a\) state (the \(+1\) eigenstate of \(X_a\)).

We can again define an associated stability matrix for the two-generator case:
\begin{equation}
\bs{\mathcal M}_{\hat C_{ab}} = \left( \begin{array}{ccccc} 1 & -1 & 0 & 0\\ 0 & 1 & 0 & 0\\ 0 & 0 & 1 & 0\\ 0 & 0 & 1 & 1\end{array}\right),
\end{equation}
Generally, 
\begin{equation}
  W_{\hat C_{ab} \ket{\alpha}}( \bs x) = W_{\ket{\alpha}}(\bs{\mathcal M}_{\hat C_{ab}} \bs x),
\end{equation}
however when \(a\) is in an \(r\)-state and \(b\) is in a \(p\)-state, or \(a\) is in a \(p\)-state and \(b\) is in a \(r\)-state, then the translation vector \(\bs r\) is also applied:
\begin{equation}
  W_{\hat C_{ab} \ket r}( \bs x) = W_{\ket r}\left(\bs{\mathcal M}_{\hat C_{ab}} \bs x + \bs r \right),
\end{equation}
where \(\bs r\) can be equivalently \((1, 0, 0, 0)\), \((0, 1, 0, 0)\), \((0, 0, 1, 0)\), or \((0, 0, 0, 1)\). This is summarized in Table~\ref{tab:cnotrvec}.
\begin{table}
  \begin{tabular}{|c|c|c|}
    \hline
    State of \(a\)& State of \(b\)& Vector \(\bs r\)\\
    \hline
    \(p\) & \(p\) & \(\bs 0\)\\
    \(q\) & \(q\) & \(\bs 0\)\\
    \(r\) & \(r\) & \(\bs 0\)\\
    \(p\) & \(q\) & \(\bs 0\)\\
    \(q\) & \(p\) & \(\bs 0\)\\
    \(p\) & \(r\) & \((1,0,0,0)\)\\
    \(r\) & \(p\) & \(\bs 0\)\\
    \(q\) & \(r\) & \(\bs 0\)\\
    \(r\) & \(q\) & \((1,0,0,0)\)\\
    \hline
  \end{tabular}
  \caption{\(\bs r\)-vector cases for the CNOT gate.}
  \label{tab:cnotrvec}
\end{table}

In the Aaronson-Gottesman's tableau algorithm, this is equivalent to the binary arithmetic ``set \(r_i := r_i \oplus x_{i_a} z_{i_b} (x_{i_b} \oplus z_{i_a} \oplus 1)\)''.

Therefore, we see that compared to the odd \(d\) case, the state to be propagated determines how the corresponding \(\bs{\mathcal M}\) matrix is used. In particular, whether the state is a \(p\) or \(q\)-state or an \(r\)-state determines whether the vector \(\bs r\) is added or not.

\subsection{Contextuality}
\label{sec:twogencontextuality}

It is thus possible to express the evolution of Clifford gates in the two-generator Wigner-Weyl-Moyal formalism to lowest order in \(\hbar^0\)---equivalent to the above \(\bs{\mathcal M}\) matrices---as long as the path integral is made dependent on the state.

Therefore, a key different aspect to describing qubit Clifford gates with the two-generator center representation is that though they still take Weyl phase space points to themselves, they don't treat Weyl phase space points independently of each other. This means that it is not possible, unlike in the odd \(d\) case, for stabilizer two-generator qubit Wigner states \(W(\bs x)\) to be evolved by \(W(\bs{\mathcal M}_{\hat O} \bs x)\) for a single symplectic \(\bs{\mathcal M}_{\hat O}\) corresponding to Clifford gate \(\hat O\). This is possible to accomplish in three-generator Weyl evolution, where arbitrary qubit stabilizer Wigner states \(\rho(\xi)\) can be evolved by \(\rho(\mathcal E_{\hat O})\) for some unitary \(\mathcal E_{\hat O}\), which can be associated with a permutation matrix \(\mathcal P_{\hat O}\).

Discrete Wigner quasi-probability distributions can be thought of as hidden variable theories, where the Weyl phase space points correspond to the hidden variables, when the quasi-probabilities are non-negative~\cite{Spekkens08}. Following this prescription, the state-dependence of the two-generator Wigner-Weyl-Moyal formalism makes its associated hidden variable theory non-local because the evolution of any hidden variable Weyl phase space point depends on the other phase space points, in particular, which of the others have positive support.

It can further be shown that this two-generator hidden variable theory implies preparation contextuality; if a mixed state \(\hat \rho\), which is a convex combination of \(x\)-, \(y\)- or \(z\)-stabilizer state basis elements \(\ket{\phi_k}\),
\begin{equation}
  \hat \rho = \sum_k P_k \ketbra{\phi_k}{\phi_k},
\end{equation}
can be transformed by some unitary operator \(\hat U\) such that
\begin{equation}
  \hat \rho = \sum_{i,j,k} P_k \hat U_{ij} \hat U^*_{ik} \ketbra{\phi_j}{\phi_k} = \sum_k \ketbra{\tilde \phi_k}{\tilde \phi_k},
\end{equation}
for some \(\ket{\tilde \phi_k}\) that are also subnormalized \(x\)-, \(y\)- and \(z\) stabilizer states, then the two ensembles require a different hidden variable theory for Clifford evolution. Despite the equality of the density operator from these two preparation schemes, the two-generator hidden variable theory for their evolution under Clifford gates is different for one compared to the other; it is thus dependent on the context.

As an example, we can consider the mixed state
\begin{widetext}
\begin{eqnarray}
  \label{eq:mixedstateex1}
  \hat \rho &=& c_{X_+}\ketbra{X_+}{X_+} + c_{X_-}\ketbra{X_-}{X_-} + c_{Z_+}\ketbra{Z_+}{Z_+} + c_{Z_-}\ketbra{Z_-}{Z_-}\\
  \label{eq:mixedstateex2}
            &=& c_{X_+}\ketbra{X_+}{X_+} + c_{X_-}\ketbra{X_-}{X_-} + c_Y \ketbra{Y_+}{Y_+} + c_Y \ketbra{Y_-}{Y_-}\\
            && \quad + (c_{Z_+}- c_Y)\ketbra{Z_+}{Z_+} + (c_{Z_-}-c_Y)\ketbra{Z_-}{Z_-}, \nonumber
\end{eqnarray}
\end{widetext}
where \(\ketbra{\alpha_\pm}{\alpha_\pm}\) denotes the projector onto the \(+1\) and \(-1\) eigenstate of \(\hat \alpha \in \{\hat X, \hat Y, \hat Z\}\) and \(c_{\alpha_\pm} \ge 0\) such that \(\Tr \hat \rho = 1\). The \(x\)- and \(z\)-basis preparation must evolve under a different two-generator hidden variable theory compared to the \(y\)-basis. Thus the preparation in the ensemble denoted by line~\ref{eq:mixedstateex1} evolves under a different hidden variable theory compared to the ensemble denoted by line~\ref{eq:mixedstateex2} for non-zero \(c_Y\).

As a result, the two-generator hidden variable theory is non-local and is (preparation-)contextual. It is a \emph{contextual} description of a \emph{non-contextual} process.

This is not the case for the three-generator formalism which we described in Section~\ref{sec:classicalityofcliffops}. There we found that the three generator Weyl phase space points evolve independently. Stabilizer state propagation therefore depends on the \emph{average} \(\hat \rho\), not on its particular realization or preparation in a basis. Therefore, the associated hidden variable theory to the three-generator formalism for qubit stabilizer state propagation under Clifford gates is local and non-contextual.

Finally, in our previous analysis of the two-generator Wigner-Weyl-Moyal formalism for odd \(d\)-dimensional qudits~\cite{Kocia16}, we found that non-contextuality was associated with the ability to treat Clifford gate evolution on stabilizer states by a finite-sum path integral truncated at \(\hbar^0\). This association also holds here for qubits. Namely, the three-generator Wigner-Weyl-Moyal formalism is able to treat Clifford gate evolution of qubit stabilizer states by a finite-term path integral truncated at order \(\hbar^0\) and can be described as a non-contextual hidden variable theory for this task. On the other hand, the two-generator Wigner-Weyl-Moyal formalism requires higher than order \(\hbar^0\) to produce a finite-term path integral describing Clifford gate evolution on stabilizer states, and it is also a contextual hidden variable theory.

\section{Pauli Measurement}
\label{sec:measurement}

We have shown so far that under the three-generator Wigner-Weyl-Moyal formalism, the Weyl symbols of stabilizer states are non-negative and can be defined over a discrete Weyl phase space. We also showed that Clifford gates are positive maps in this formalism and can be formulated in terms of permutation matrices acting on the discrete Weyl phase space, such that they take non-negative states to other non-negative states.

What remains to complete the Clifford operations are measurements in the Pauli basis. Unlike the preparation and unitary propagation part of Clifford operations, which are manifestly non-contextual from the point of view of preparation contextuality, Clifford measurements can be contextual (from the point of view of measurement contextuality). A well-known example of this is demonstrated by the Peres-Mermin square~\cite{Peres90,Mermin90} shown in Table~\ref{tab:peresmermin}. Here we show that the three-generator Wigner-Weyl-Moyal formalism is contextual for Pauli measurements in the Peres-Mermin square because its Weyl symbols for measurement operators \(\Pi\) produce expectation values depending on the measurement context that cannot be represented by the average of an indicator function, \(\mathcal I_\Pi \in [0,1]\), with the previously defined associated probability distributions \(\bar g\) for stabilizer states.

The Weyl symbols (\(\xi_r \xi_q\), \(\xi_p \xi_q\), \(\xi_p \xi_r\)) of single qubit Pauli (\(\hat \sigma_p\), \(\hat \sigma_r\), \(\hat \sigma_q\)) observables are positive maps; they take Weyl phase space points to themselves. The Weyl symbols of the projection operators onto their eigenstates are also positive. Any multiqubit Clifford measurement can be reexpressed as a sequence of Clifford gates and then a single qubit measurement. Therefore, every step has an associated Weyl symbol that is non-negative and so the Weyl symbols of multiqubit Pauli measurements take stabilizer states to themselves.

Nevertheless, the three-generator Wigner-Weyl-Moyal formalism is contextual for the Peres-Mermin measurements in a very similar way that the two-generator Wigner-Weyl-Moyal formalism was contextual for unitary qubit Clifford gates. We showed that the two-generator formalism, also described by the Aaronson-Gottesman tableau algorithm, has positive but contextual transformation of qubit stabilizer states under Clifford gates because its hidden variable theory depends on the state or preparation context. In the three-generator formalism, we will show measurement contextuality by demonstrating that the associated hidden variable theory depends on the measurement context. In particular, we will show that the context of the measurements changes the associated three-generator Weyl symbol's expectation values for the measurements (and therefore any associated permutation matrix). In this way, the three-generator Wigner-Weyl-Moyal formalism describes a \emph{contextual} measurement process \emph{contextually}. Therefore, the three generator Wigner-Weyl-Moyal formalism has a positive but contextual transformation~\cite{Liang11} of stabilizer staes under Pauli measurements.

\subsection{Peres-Mermin Square}
\label{sec:peresmermin}

\begin{table}
  \begin{tabular}{c|c|c|c||c}
    & Meas. \# 1 & Meas. \# 2 & Meas. \# 3 & Outcome\\
    \hline
    Meas. \# 1 & \(\hat \sigma_{p_1}\) & \(\hat \sigma_{p_2}\) & \(\hat \sigma_{p_1} \hat \sigma_{p_2}\) & \(+1\)\\
    \hline
    Meas. \# 2 & \(\hat \sigma_{r_2}\) & \(\hat \sigma_{r_1}\) & \(\hat \sigma_{r_1} \hat \sigma_{r_2}\) & \(+1\)\\
    \hline
    Meas. \# 3 & \(\hat \sigma_{p_1} \hat \sigma_{r_2}\) & \(\hat \sigma_{r_1} \hat \sigma_{p_2}\) & \(\hat \sigma_{q_1} \hat \sigma_{q_2}\) & \(+1\)\\
    \hline
    \hline
    Outcome & \(+1\) & \(+1\) & \(-1\) & \backslashbox{\(-1\)}{\(+1\)}
  \end{tabular}
  \caption{The Peres-Mermin Square~\cite{Peres90,Mermin90}. Every observable commutes with every other observable in its row and column, but anticommutes with the other four observables. Taking the measurements row-wise produces only \(+1\) outcomes (by Eq.~\ref{eq:Paulimatrixrelation}), while the measurements column-wise produce two \(+1\) outcomes and a \(-1\) outcome, the product of which is \(-1\) as shown in the bottom-rightmost cell. Hence, the context of the measurement scheme determines the outcomes.}
  \label{tab:peresmermin}
\end{table}

Measurement contextuality can be seen in the Peres-Mermin square (see Table~\ref{tab:peresmermin}) where every entry contains a two-qubit Pauli measurement of the form \(\hat \sigma_\alpha \hat \sigma_\beta\) for \(\alpha, \beta \in \{p_1, p_2, q_1, q_2, r_1, r_2\}\), which has the associated Weyl symbol \(\frac{1}{4} \sum_{j k} \epsilon_{\alpha j k} \xi_j \xi_k \sum_{l m} \epsilon_{\beta l m} \xi_l \xi_m\). Every observable in the table commutes with the other observables in its row and column and anticommutes with the other four observables in the table. Thus we can make all the measurements in the Peres-Mermin square row-wise or column-wise and compare the results obtained.

The first row of measurements \(\hat \sigma_{p_1} \), \(\hat \sigma_{p_2} \), and \(\hat \sigma_{p_1} \hat \sigma_{p_2}\) are associated with the projector-valued measurements \(\hat \Pi_{11}^{(m_{11})}\), \(\hat \Pi_{12}^{(m_{12})}\) and \(\hat \Pi_{13}^{(m_{13})}\) for \(m_{11}, m_{12}, m_{13} \in \{+1, -1\}\). These have the associated projectors,
\begin{eqnarray}
  \hat \Pi_{11}^{+} &=& \ketbra{+}{+} \otimes \mathbb I, \nonumber\\
  \hat \Pi_{11}^{-} &=& \ketbra{-}{-} \otimes \mathbb I, \nonumber\\
  \hat \Pi_{12}^{+} &=& \mathbb I \otimes \ketbra{+}{+}, \nonumber\\
  \hat \Pi_{12}^{-} &=& \mathbb I \otimes \ketbra{-}{-},\\
  \hat \Pi_{22}^{+} &=& ( \ketbra{++}{++} + \ketbra{--}{--} ), \nonumber\\
  \hat \Pi_{22}^{-} &=& (\ketbra{+-}{+-} + \ketbra{-+}{-+}). \nonumber
\end{eqnarray}
These projectors have associated Weyl symbols:
\begin{eqnarray}
  \Pi_{11}^{+}(\bs \xi_1, \bs \xi_2) &=& \frac{1}{2}(1 + i \xi_{r_1} \xi_{p_1}), \nonumber\\
  \label{eq:measurementPVMgrassmann1}
  \Pi_{11}^{-}(\bs \xi_1, \bs \xi_2) &=& \frac{1}{2}(1 - i \xi_{r_1} \xi_{p_1}), \nonumber\\
  \Pi_{12}^{+}(\bs \xi_1, \bs \xi_2) &=& \frac{1}{2}(1 + i \xi_{r_2} \xi_{p_2}), \nonumber\\
  \Pi_{12}^{-}(\bs \xi_1, \bs \xi_2) &=& \frac{1}{2}(1 - i \xi_{r_2} \xi_{p_2}), \\
  \Pi_{22}^{+}(\bs \xi_1, \bs \xi_2) &=& \frac{1}{4}(1 + i \xi_{r_1} \xi_{p_1})(1 + i \xi_{r_2} \xi_{p_2})  \nonumber\\
                        && + \frac{1}{4}(1 - i \xi_{r_1} \xi_{p_1})(1 - i \xi_{r_2} \xi_{p_2}), \nonumber\\
  \Pi_{22}^{-}(\bs \xi_1, \bs \xi_2) &=& \frac{1}{4}(1 + i \xi_{r_1} \xi_{p_1})(1 - i \xi_{r_2} \xi_{p_2}) \nonumber\\
                        && + \frac{1}{4}(1 - i \xi_{r_1} \xi_{p_1})(1 + i \xi_{r_2} \xi_{p_2}). \nonumber
\end{eqnarray}
The outcome of the row's measurements is determined from just two \(\Pi^\pm_{ij}\) projector-value measures (since the product of outcomes in the row must equal \(+1\)), and choosing any pair of \(\Pi^\pm_{ij}\) is equivalent to choosing the context of the measurement~\cite{Spekkens08}.

There is a single measurement that simulates these two compatible measurements. This is the measurement with the projectors:
\begin{eqnarray}
  \Pi_{R_1}^{(m_{R_{1,1}}, m_{R_{1,2}})} &=& \frac{1}{4} \left( \mathbb I \otimes \mathbb I + m_{R_{1,1}} \hat \sigma_{p_1} \otimes \mathbb I\right.\\
  && \left.+ m_{R_{1,2}} \mathbb I \otimes \hat \sigma_{p_2} + m_{R_{1,1}} m_{R_{1,2}} \hat \sigma_{p_1} \hat \sigma_{p_2} \right). \nonumber
\end{eqnarray}
The associated projectors are thus
\begin{equation}
  \Pi_{R_1}^{(m_{{R_1},1}, m_{{R_1},2})} = \ketbra{m_{{R_1},1}, m_{{R_1},2}}{m_{{R_1},1}, m_{{R_1},2}}
\end{equation}
for \(m_{{R_1},1}, m_{{R_1},2} \in \{+1, -1\}\). These have associated Weyl symbols
\begin{eqnarray}
  \label{eq:measurementPVMgrassmann2}
  \Pi_{R_1}^{(m_{{R_1},1}, m_{{R_1},2})}(\bs \xi_1, \bs \xi_2) &=& \frac{1}{4}(1-i m_{{R_1},1} \xi_{r_1}\xi_{q_1})\\
                                                               &&\times (1-i m_{{R_1},2} \xi_{r_2}\xi_{q_2}).  \nonumber
\end{eqnarray}
As a result, if for instance one chooses to measure in the context corresponding to the projector-valued measurements \(\Pi_{11}^\pm(\bs \xi)\) and \(\Pi_{12}^\pm(\bs \xi)\), the product of their expectation values \(\int \Pi_{11}^+(\bs \xi_1, \bs \xi_2) \rho(\bs \xi_1, \bs \xi_2) \text d^3 \bs \xi_1 \text d^3 \bs \xi_2 \) and \(\int \Pi_{12}^+(\bs \xi_1, \bs \xi_2) \rho(\bs \xi_1 \bs \xi_2) \text d^3 \bs \xi_1 \text d^3 \bs \xi_2 \) is conditioned to be equal to the expectation value of \(\int \Pi_{R_1}^{(+1, +1)}(\bs \xi_1, \bs \xi_2) \rho(\bs \xi_1, \bs \xi_2) \text d^3 \bs \xi_1 \text d^3 \bs \xi_2 \). This means, for instance, that the hidden variable predicting the expectation values of \(\Pi^\pm_{12}\) is different depending on whether it is measured in context with \(\Pi^\pm_{11}\) or with \(\Pi^\pm_{12}\). Since the three-generator formalism is a faithful representation of quantum mechanics, it satisfies these conditions, as can be readily verified. It follows that the three-generator formalism for that row's measurement operators is contextual by the operational description of the Peres-Mermin square~\cite{Krishna17}. The same can be found for the other two rows and the columns.

In this way, the three-generator Wigner-Weyl-Moyal formalism indicates measurement-contextuality by producing different expectation values for the Weyl symbols of the measurement operators depending on the context of the measurement scheme. In other words, it is not possible to associate a real valued indicator function \(\mathcal I_{\Pi_{11}}, \mathcal I_{\Pi_{12}}, \mathcal I_{\Pi_{22}} \in [0,1]\) that reproduces these results with the associated \(\bar g\) distributions for stabilizer states in Eq.~\ref{eq:probdistrogbar} by \(\int_\Lambda \bar g(\lambda) \mathcal I_{\Pi_{ij}}(\lambda) \text d \lambda\) for \(i, j \in \{1,2\}\)~\cite{Spekkens08}.

We note however, that systems comprised of a single qubit stabilizer state, Clifford gates (which necessarily don't include the two-qubit CNOT gate), and Pauli measurements on this single qubit, are non-contextual as shown by Wallman \emph{et al}.~\cite{Wallman12}, as such indicator functions \(\mathcal I\) do exists with three generators.

\section{Conclusion}

Contextuality has been shown to be a necessary resource for universal quantum computation via magic state distillation for qudits of any odd dimension~\cite{Howard14}. The same result has been recently proposed for qubits that satisfy additional postulates~\cite{Raussendorf15}. Furthermore, non-contextuality has been shown to be equivalent to the non-negativity of the discrete Wigner functions for odd \(d\)-dimensional qudits~\cite{Gross08,Ferrie08,Ferrie09,Veitch12,Mari12}. We have extended these results to \(d=2\) and shown that Clifford gates on qubit stabilizer states are non-contextual and that their appropriate Weyl symbols have associated non-negative probability distributions. On the other hand, we showed the Pauli measurements are contextual and different measurement contexts produce different Weyl symbols with associated expectation values that are appropriately contextual.

To demonstrate non-contextuality for qubit stabilizer states under Clifford gates, we relied on three generators, \(\hat p\), \(\hat q\) and \(\hat r\), instead of the usual two, \(\hat p\) and \(\hat q\), to produce the Wigner-Weyl-Moyal formalism that defines our Weyl symbols. This was necessary because the Weyl-Heisenberg \(\hat T\)-operator group that forms a basis for the two-generator formalism is not a subgroup of \(SU(d)\) for even \(d\). Equivalently, since the Clifford gates are a three-design, it is not possible to express their action on the \(\hat R\)-operator basis, which is dual to the Weyl-Heisenberg \(\hat T\)-operator basis, in a covariant manner~\cite{Zhu15}.

We showed that the resultant three-generator Weyl symbols of the stabilizer states have associated non-negative probability distributions whose evolution under Clifford gates can be described by a local and non-contextual hidden-variable theory. It was further shown that the three-generator Wigner-Weyl-Moyal formalism produces Weyl propagators for the Clifford gates that can be truncated to order \(\hbar^0\) with a finite number of terms with no loss of information. On the other hand, \(T\)-gates where found to require Weyl propagators that were expanded up to order \(\hbar^1\), and Pauli measurements were found to produce contextual three-generator Weyl symbols.

We showed that employing a two-generator Wigner-Weyl-Moyal formalism, as has been done for odd \(d\)-dimensional qudits~\cite{Kocia16,Veitch12,Mari12,Gross08,Almeida98}, produces a non-local and contextual description of Clifford gates on qubits. This produces state-dependent evolution and explains the Aaronson-Gottesman tableau algorithm's unitary evolution rules. In other words, the two-generator Wigner-Weyl-Moyal formalism produces a \emph{contextual} description of the \emph{non-contextual} Clifford gate evolution process. Equivalently, the two-generator Wigner-Weyl-Moyal Clifford symbols require a treatment at order \(\hbar^1\) to describe evolution that is possible at order \(\hbar^0\) by three-generator Weyl symbols.

In summary, this paper shows that the classical nature of Clifford gates on stabilizer states is likely well characterized for all \(d\)-dimensional qudits; it is a non-contextual process that can be described by a local hidden variable theory. An example of such a hidden variable theory involves treating Weyl phase space as the hidden variables that evolve independently under classical harmonic Hamiltonians. For such hidden variable theories, we have shown that there always exists an appropriately defined Wigner-Weyl-Moyal formalism that produces discrete stabilizer state Wigner functions with associated probability distributions that are non-negative. Clifford gates on stabilizer states are just the discrete analog of harmonic evolution of Gaussian states in the continuous case, and are thus fully treatable by path integrals at order \(\hbar^0\) (the classical limit). On the other hand, the same appropriate Wigner-Weyl-Moyal formalism shows that Pauli measurements, which complete the set of allowed Clifford operations, can introduce contextuality into a scheme.

\section{Acknowledgments}
The authors thank Yifei Huang, Byron Drury, Prof. Joseph Emerson and Prof. Alfredo Ozorio de Almeida. This work was supported by AFOSR award no. FA9550-12-1-0046.

\appendix
\section*{Appendix}
\label{sec:appendix}

We can construct a matrix representation of the Grassmann generators from the corresponding Clifford algebra. Consider the generators \(\xi_k\) of left multiplication by \(\xi_k\) and \(\partder{}{\xi_k}\) of left differentiation by \(\xi_k\), which obey
\begin{equation}
  \left\{\xi_k, \partder{}{\xi_j}\right\} = \delta_{kj}.
\end{equation}
We can now form the operators \(q_k = \xi_k + \partder{}{\xi_k}\) and \(p_k = -i(\xi_k - \partder{}{\xi_k})\) of the Clifford algebra that corresponds to this Grassmann algebra~\cite{Berezin66}. These operators satisfy the relations
\begin{eqnarray}
  \{p_i, q_j\} &=& 0,\\
  \{p_i, p_j\} = \{q_i,q_j\} &=& 2 \delta_{ij}.
\end{eqnarray}

Each Clifford subalgebra generated by the three \(p_k\) and \(q_k\) can be represented by the eight by eight matrices with complex entries defined over the field \(\mathbb C\)~\cite{Traubenberg05,Catto13}
\begin{eqnarray}
  \xi_p &=& \frac{1}{2}(q_p + i p_p) \mapsto \hat \sigma_x \otimes \hat I \otimes \hat I + i \hat \sigma_y \otimes \hat I \otimes \hat I,\\
  \xi_r &=& \frac{1}{2}(q_r + i p_r) \mapsto \hat \sigma_z \otimes \hat \sigma_x \otimes \hat I + i \hat \sigma_z \otimes \hat \sigma_y \otimes \hat I,\\
  \xi_q &=& \frac{1}{2}(q_q + i p_q) \mapsto \hat \sigma_z \otimes \hat \sigma_z \otimes \hat \sigma_x + i \hat \sigma_z \otimes \hat \sigma_z \otimes \hat \sigma_y,
\end{eqnarray}
where \(\mapsto\) denotes a representation (an algebra homomorphism) and \(\otimes\) denotes a matrix outer product.
One can verify that
\begin{equation}
  \xi_k^2 = (\hat \sigma_x^2 +i \{\hat \sigma_x \hat \sigma_y\} - \hat \sigma_y^2) = ( \hat I - \hat I) = 0
\end{equation}
for all \(k\) as expected.

With these matrices in hand, we can see that a Weyl symbol \(g(\xi)\) is really just a a representation of the operator \(\hat g\), which is in the Clifford algebra, in a higher-dimensional Grassmann algebra. However, the latter Grassmann algebra is generated by elements whose evolution is governed by a Poisson bracket, and therefore function like classical conjugate degrees of freedom, though instead of commuting with each other, they anti-commute.

\bibliography{biblio}{}
\bibliographystyle{unsrt}

\end{document}